\input harvmac
\input amssym

\def\unit{\relax{\rm 1\kern-.26em I}}
\def\nada{\relax{\rm 0\kern-.30em l}}
\def\tilde{\widetilde}



\def\det{{\rm det}}

\noblackbox
\def\IL{\relax{\rm I\kern-.18em L}}
\def\IH{\relax{\rm I\kern-.18em H}}
\def\IR{\relax{\rm I\kern-.18em R}}
\def\IC{\relax\hbox{$\inbar\kern-.3em{\rm C}$}}
\def\IZ{\relax\ifmmode\mathchoice
{\hbox{\cmss Z\kern-.4em Z}}{\hbox{\cmss Z\kern-.4em Z}}
{\lower.9pt\hbox{\cmsss Z\kern-.4em Z}} {\lower1.2pt\hbox{\cmsss
Z\kern-.4em Z}}\else{\cmss Z\kern-.4em Z}\fi}

\def\CN {{\cal N}}
\def\CR {{\cal R}}

\def\CO {{\cal O}}


\def\CN {{\cal N}}

\def\CO {{\cal O}}

\def\CQ {{\cal Q }}

\def\det{{\rm det}}
\def\Tr{{\rm Tr}}

\font\manual=manfnt \def\dbend{\lower3.5pt\hbox{\manual\char127}}

\def\IZ{\relax\ifmmode\mathchoice
{\hbox{\cmss Z\kern-.4em Z}}{\hbox{\cmss Z\kern-.4em Z}}
{\lower.9pt\hbox{\cmsss Z\kern-.4em Z}} {\lower1.2pt\hbox{\cmsss
Z\kern-.4em Z}}\else{\cmss Z\kern-.4em Z}\fi}

\def\bar{\overline}

\def\rt2{\sqrt{2}}
\def\irt2{{1\over\sqrt{2}}}

\def\hat{\widehat}
\def\slashchar#1{\setbox0=\hbox{$#1$}           
   \dimen0=\wd0                                 
   \setbox1=\hbox{/} \dimen1=\wd1               
   \ifdim\dimen0>\dimen1                        
      \rlap{\hbox to \dimen0{\hfil/\hfil}}      
      #1                                        
   \else                                        
      \rlap{\hbox to \dimen1{\hfil$#1$\hfil}}   
      /                                         
   \fi}

\def\foursqr#1#2{{\vcenter{\vbox{
    \hrule height.#2pt
    \hbox{\vrule width.#2pt height#1pt \kern#1pt
    \vrule width.#2pt}
    \hrule height.#2pt
    \hrule height.#2pt
    \hbox{\vrule width.#2pt height#1pt \kern#1pt
    \vrule width.#2pt}
    \hrule height.#2pt
        \hrule height.#2pt
    \hbox{\vrule width.#2pt height#1pt \kern#1pt
    \vrule width.#2pt}
    \hrule height.#2pt
        \hrule height.#2pt
    \hbox{\vrule width.#2pt height#1pt \kern#1pt
    \vrule width.#2pt}
    \hrule height.#2pt}}}}
\def\psqr#1#2{{\vcenter{\vbox{\hrule height.#2pt
    \hbox{\vrule width.#2pt height#1pt \kern#1pt
    \vrule width.#2pt}
    \hrule height.#2pt \hrule height.#2pt
    \hbox{\vrule width.#2pt height#1pt \kern#1pt
    \vrule width.#2pt}
    \hrule height.#2pt}}}}
\def\sqr#1#2{{\vcenter{\vbox{\hrule height.#2pt
    \hbox{\vrule width.#2pt height#1pt \kern#1pt
    \vrule width.#2pt}
    \hrule height.#2pt}}}}

\def\figin{\epsfcheck\figin}\def\figins{\epsfcheck\figins}
\def\epsfcheck{\ifx\epsfbox\UnDeFiNeD
\message{(NO epsf.tex, FIGURES WILL BE IGNORED)}
\gdef\figin##1{\vskip2in}\gdef\figins##1{\hskip.5in}
\else\message{(FIGURES WILL BE INCLUDED)}%
\gdef\figin##1{##1}\gdef\figins##1{##1}\fi}
\def\DefWarn#1{}
\def\figinsert{\goodbreak\midinsert}
\def\ifig#1#2#3{\DefWarn#1\xdef#1{fig.~\the\figno}
\writedef{#1\leftbracket fig.\noexpand~\the\figno}%
\figinsert\figin{\centerline{#3}}\medskip\centerline{\vbox{\baselineskip12pt
\advance\hsize by -1truein\noindent\footnotefont{\bf
Fig.~\the\figno:\ } \it#2}}
\bigskip\endinsert\global\advance\figno by1}

\lref\KutasovIY{
  D.~Kutasov, A.~Parnachev, D.~A.~Sahakyan,
  ``Central charges and U(1)(R) symmetries in N=1 superYang-Mills,''
JHEP {\bf 0311}, 013 (2003).
[hep-th/0308071].
}

\lref\DorigoniRA{
  D.~Dorigoni, V.~S.~Rychkov,
  ``Scale Invariance + Unitarity $=>$ Conformal Invariance?,''
[arXiv:0910.1087 [hep-th]].
}

\lref\IntriligatorJJ{
  K.~A.~Intriligator, B.~Wecht,
  ``The Exact superconformal R symmetry maximizes a,''
Nucl.\ Phys.\  {\bf B667}, 183-200 (2003).
[hep-th/0304128].
}

\lref\BarnesJJ{
  E.~Barnes, K.~A.~Intriligator, B.~Wecht, J.~Wright,
  ``Evidence for the strongest version of the 4d a-theorem, via a-maximization along RG flows,''
Nucl.\ Phys.\  {\bf B702}, 131-162 (2004).
[hep-th/0408156].
}

\lref\KutasovUX{
  D.~Kutasov,
  ``New results on the 'a theorem' in four-dimensional supersymmetric field theory,''
[hep-th/0312098].
}

\lref\KutasovXU{
  D.~Kutasov, A.~Schwimmer,
  ``Lagrange multipliers and couplings in supersymmetric field theory,''
Nucl.\ Phys.\  {\bf B702}, 369-379 (2004).
[hep-th/0409029].
}

\lref\AnselmiAM{
  D.~Anselmi, D.~Z.~Freedman, M.~T.~Grisaru, A.~A.~Johansen,
  ``Nonperturbative formulas for central functions of supersymmetric gauge theories,''
Nucl.\ Phys.\  {\bf B526}, 543-571 (1998).
[hep-th/9708042].
}

\lref\AnselmiYS{
  D.~Anselmi, J.~Erlich, D.~Z.~Freedman, A.~A.~Johansen,
  ``Positivity constraints on anomalies in supersymmetric gauge theories,''
Phys.\ Rev.\  {\bf D57}, 7570-7588 (1998).
[hep-th/9711035].
}

\lref\FerraraPZ{
  S.~Ferrara, B.~Zumino,
  ``Transformation Properties of the Supercurrent,''
Nucl.\ Phys.\  {\bf B87}, 207 (1975).
}

\lref\KomargodskiRB{
  Z.~Komargodski, N.~Seiberg,
  ``Comments on Supercurrent Multiplets, Supersymmetric Field Theories and Supergravity,''
JHEP {\bf 1007}, 017 (2010).
[arXiv:1002.2228 [hep-th]].
}

\lref\IntriligatorIF{
  K.~A.~Intriligator,
  ``IR free or interacting? A Proposed diagnostic,''
Nucl.\ Phys.\  {\bf B730}, 239-251 (2005).
[hep-th/0509085].
}

\lref\CappelliYC{
  A.~Cappelli, D.~Friedan, J.~I.~Latorre,
  ``C theorem and spectral representation,''
Nucl.\ Phys.\  {\bf B352}, 616-670 (1991).
}

\lref\AppelquistHR{
  T.~Appelquist, A.~G.~Cohen, M.~Schmaltz,
  ``A New constraint on strongly coupled gauge theories,''
Phys.\ Rev.\  {\bf D60}, 045003 (1999).
[arXiv:hep-th/9901109 [hep-th]].
}

\lref\AbelWV{
  S.~Abel, M.~Buican, Z.~Komargodski,
  ``Mapping Anomalous Currents in Supersymmetric Dualities,''
[arXiv:1105.2885 [hep-th]].
}

\lref\BrodieVX{
  J.~H.~Brodie,
  ``Duality in supersymmetric SU(N(c)) gauge theory with two adjoint chiral superfields,''
Nucl.\ Phys.\  {\bf B478}, 123-140 (1996).
[hep-th/9605232].
}

\lref\IntriligatorMI{
  K.~A.~Intriligator, B.~Wecht,
  ``RG fixed points and flows in SQCD with adjoints,''
Nucl.\ Phys.\  {\bf B677}, 223-272 (2004).
[hep-th/0309201].
}

\lref\LeighDS{
  R.~G.~Leigh, M.~J.~Strassler,
  ``Accidental symmetries and N=1 duality in supersymmetric gauge theory,''
Nucl.\ Phys.\  {\bf B496}, 132-148 (1997).
[hep-th/9611020].
}

\lref\PolandWG{
  D.~Poland, D.~Simmons-Duffin,
  ``Bounds on 4D Conformal and Superconformal Field Theories,''
JHEP {\bf 1105}, 017 (2011).
[arXiv:1009.2087 [hep-th]].
}

\lref\BarnesBM{
  E.~Barnes, E.~Gorbatov, K.~A.~Intriligator, M.~Sudano, J.~Wright,
  ``The Exact superconformal R-symmetry minimizes tau(RR),''
Nucl.\ Phys.\  {\bf B730}, 210-222 (2005).
[hep-th/0507137].
}

\lref\HofmanAR{
  D.~M.~Hofman, J.~Maldacena,
  ``Conformal collider physics: Energy and charge correlations,''
JHEP {\bf 0805}, 012 (2008).
[arXiv:0803.1467 [hep-th]].
}

\lref\CsakiZB{
  C.~Csaki, M.~Schmaltz, W.~Skiba,
  ``Confinement in N=1 SUSY gauge theories and model building tools,''
Phys.\ Rev.\  {\bf D55}, 7840-7858 (1997).
[hep-th/9612207].
}

\lref\SeibergRS{
  N.~Seiberg, E.~Witten,
  ``Electric - magnetic duality, monopole condensation, and confinement in N=2 supersymmetric Yang-Mills theory,''
Nucl.\ Phys.\  {\bf B426}, 19-52 (1994).
[hep-th/9407087].
}

\lref\DouglasNW{
  M.~R.~Douglas, S.~H.~Shenker,
  ``Dynamics of SU(N) supersymmetric gauge theory,''
Nucl.\ Phys.\  {\bf B447}, 271-296 (1995).
[hep-th/9503163].
}

\lref\FortinKS{
  J.~-F.~Fortin, B.~Grinstein, A.~Stergiou,
  ``Scale without Conformal Invariance: An Example,''
[arXiv:1106.2540 [hep-th]].
}

\lref\AntoniadisGN{
  I.~Antoniadis, M.~Buican,
  ``On R-symmetric Fixed Points and Superconformality,''
Phys.\ Rev.\  {\bf D83}, 105011 (2011).
[arXiv:1102.2294 [hep-th]].
}

\lref\ZamolodchikovGT{
  A.~B.~Zamolodchikov,
  ``Irreversibility of the Flux of the Renormalization Group in a 2D Field Theory,''
JETP Lett.\  {\bf 43}, 730-732 (1986).
}

\lref\PolchinskiDY{
  J.~Polchinski,
  ``Scale And Conformal Invariance In Quantum Field Theory,''
Nucl.\ Phys.\  {\bf B303}, 226 (1988).
}

\lref\CardyCWA{
  J.~L.~Cardy,
  ``Is There a c Theorem in Four-Dimensions?,''
Phys.\ Lett.\  {\bf B215}, 749-752 (1988).
}

\lref\SeibergPQ{
  N.~Seiberg,
  ``Electric - magnetic duality in supersymmetric nonAbelian gauge theories,''
Nucl.\ Phys.\  {\bf B435}, 129-146 (1995).
[hep-th/9411149].
}

\lref\KomargodskiVJ{
  Z.~Komargodski and A.~Schwimmer,
  ``On Renormalization Group Flows in Four Dimensions,''
  arXiv:1107.3987 [hep-th].
}

\lref\IntriligatorID{
  K.~A.~Intriligator, N.~Seiberg,
  ``Duality, monopoles, dyons, confinement and oblique confinement in supersymmetric SO(N(c)) gauge theories,''
Nucl.\ Phys.\  {\bf B444}, 125-160 (1995).
[hep-th/9503179].
}

\lref\IntriligatorAU{
  K.~A.~Intriligator, N.~Seiberg,
  ``Lectures on supersymmetric gauge theories and electric - magnetic duality,''
Nucl.\ Phys.\ Proc.\ Suppl.\  {\bf 45BC}, 1-28 (1996).
[hep-th/9509066].
}

\lref\IntriligatorNE{
  K.~A.~Intriligator, P.~Pouliot,
  ``Exact superpotentials, quantum vacua and duality in supersymmetric SP(N(c)) gauge theories,''
Phys.\ Lett.\  {\bf B353}, 471-476 (1995).
[hep-th/9505006].
}

\lref\KutasovVE{
  D.~Kutasov,
  ``A Comment on duality in N=1 supersymmetric nonAbelian gauge theories,''
Phys.\ Lett.\  {\bf B351}, 230-234 (1995).
[hep-th/9503086].
}

\lref\KutasovSS{
  D.~Kutasov, A.~Schwimmer, N.~Seiberg,
  ``Chiral rings, singularity theory and electric - magnetic duality,''
Nucl.\ Phys.\  {\bf B459}, 455-496 (1996).
[hep-th/9510222].
}

\lref\SeibergRS{
  N.~Seiberg, E.~Witten,
  ``Electric - magnetic duality, monopole condensation, and confinement in N=2 supersymmetric Yang-Mills theory,''
Nucl.\ Phys.\  {\bf B426}, 19-52 (1994).
[hep-th/9407087].
}
\lref\DouglasNW{
  M.~R.~Douglas, S.~H.~Shenker,
  ``Dynamics of SU(N) supersymmetric gauge theory,''
Nucl.\ Phys.\  {\bf B447}, 271-296 (1995).
[hep-th/9503163].
}

\lref\ArgyresXH{
  P.~C.~Argyres, A.~E.~Faraggi,
  ``The vacuum structure and spectrum of N=2 supersymmetric SU(n) gauge theory,''
Phys.\ Rev.\ Lett.\  {\bf 74}, 3931-3934 (1995).
[hep-th/9411057].
}

\lref\KlemmQS{
  A.~Klemm, W.~Lerche, S.~Yankielowicz, S.~Theisen,
  ``Simple singularities and N=2 supersymmetric Yang-Mills theory,''
Phys.\ Lett.\  {\bf B344}, 169-175 (1995).
[hep-th/9411048].
}

\lref\KlemmQJ{
  A.~Klemm, W.~Lerche, S.~Yankielowicz, S.~Theisen,
  ``On the monodromies of N=2 supersymmetric Yang-Mills theory,''
[hep-th/9412158].
}

\lref\CsakiZB{
  C.~Csaki, M.~Schmaltz, W.~Skiba,
  ``Confinement in N=1 SUSY gauge theories and model building tools,''
Phys.\ Rev.\  {\bf D55}, 7840-7858 (1997).
[hep-th/9612207].
}

\lref\IntriligatorRX{
  K.~A.~Intriligator, N.~Seiberg, S.~H.~Shenker,
  ``Proposal for a simple model of dynamical SUSY breaking,''
Phys.\ Lett.\  {\bf B342}, 152-154 (1995).
[hep-ph/9410203].
}

\lref\IntriligatorIF{
  K.~A.~Intriligator,
  ``IR free or interacting? A Proposed diagnostic,''
Nucl.\ Phys.\  {\bf B730}, 239-251 (2005).
[hep-th/0509085].
}

\lref\VartanovXJ{
  G.~S.~Vartanov,
  ``On the ISS model of dynamical SUSY breaking,''
Phys.\ Lett.\  {\bf B696}, 288-290 (2011).
[arXiv:1009.2153 [hep-th]].
}

\lref\BrodieVV{
  J.~H.~Brodie, P.~L.~Cho, K.~A.~Intriligator,
  ``Misleading anomaly matchings?,''
Phys.\ Lett.\  {\bf B429}, 319-326 (1998).
[hep-th/9802092].
}

\lref\MyersTJ{
  R.~C.~Myers, A.~Sinha,
  ``Holographic c-theorems in arbitrary dimensions,''
JHEP {\bf 1101}, 125 (2011).
[arXiv:1011.5819 [hep-th]].
}

\lref\FreedmanGP{
  D.~Z.~Freedman, S.~S.~Gubser, K.~Pilch, N.~P.~Warner,
  ``Renormalization group flows from holography supersymmetry and a c theorem,''
Adv.\ Theor.\ Math.\ Phys.\  {\bf 3}, 363-417 (1999).
[hep-th/9904017].
}

\lref\GirardelloBD{
  L.~Girardello, M.~Petrini, M.~Porrati, A.~Zaffaroni,
  ``The Supergravity dual of N=1 superYang-Mills theory,''
Nucl.\ Phys.\  {\bf B569}, 451-469 (2000).
[hep-th/9909047].
}

\lref\JafferisUN{
  D.~L.~Jafferis,
  ``The Exact Superconformal R-Symmetry Extremizes Z,''
[arXiv:1012.3210 [hep-th]].
}

\lref\DumitrescuIU{
  T.~T.~Dumitrescu, N.~Seiberg,
  ``Supercurrents and Brane Currents in Diverse Dimensions,''
JHEP {\bf 1107}, 095 (2011).
[arXiv:1106.0031 [hep-th]].
}

\lref\DumitrescuCA{
  T.~T.~Dumitrescu, Z.~Komargodski, M.~Sudano,
  ``Global Symmetries and D-Terms in Supersymmetric Field Theories,''
JHEP {\bf 1011}, 052 (2010).
[arXiv:1007.5352 [hep-th]].
}

\lref\BarnesBM{
  E.~Barnes, E.~Gorbatov, K.~A.~Intriligator, M.~Sudano, J.~Wright,
  ``The Exact superconformal R-symmetry minimizes tau(RR),''
Nucl.\ Phys.\  {\bf B730}, 210-222 (2005).
[hep-th/0507137].
}

\lref\VichiUX{
  A.~Vichi,
  ``Improved bounds for CFT's with global symmetries,''
[arXiv:1106.4037 [hep-th]].
}

\lref\LeighDS{
  R.~G.~Leigh, M.~J.~Strassler,
  ``Accidental symmetries and N=1 duality in supersymmetric gauge theory,''
Nucl.\ Phys.\  {\bf B496}, 132-148 (1997).
[hep-th/9611020].
}

\lref\KomargodskiPC{
  Z.~Komargodski, N.~Seiberg,
  ``Comments on the Fayet-Iliopoulos Term in Field Theory and Supergravity,''
JHEP {\bf 0906}, 007 (2009).
[arXiv:0904.1159 [hep-th]].
}

\lref\PoppitzKZ{
  E.~Poppitz, M.~Unsal,
  ``Chiral gauge dynamics and dynamical supersymmetry breaking,''
JHEP {\bf 0907}, 060 (2009).
[arXiv:0905.0634 [hep-th]].
}

\lref\LutyQC{
  M.~A.~Luty, R.~Rattazzi,
  ``Soft supersymmetry breaking in deformed moduli spaces, conformal theories, and N=2 Yang-Mills theory,''
JHEP {\bf 9911}, 001 (1999).
[hep-th/9908085].
}

\lref\NakayamaWX{
  Y.~Nakayama,
  ``Higher derivative corrections in holographic Zamolodchikov-Polchinski theorem,''
[arXiv:1009.0491 [hep-th]].
}

\rightline{CERN-PH-TH/2011-223}
\Title{\vbox{\baselineskip12pt }} {\vbox{\centerline{A Conjectured Bound on Accidental Symmetries}}}
\smallskip
\centerline{Matthew Buican}
\smallskip
\bigskip
\centerline{{\it Department of Physics, CERN Theory Division, CH-1211 Geneva 23, Switzerland}} %
\vskip 1cm

\noindent
In this note, we study a large class of four-dimensional $R$-symmetric theories, and we describe a new quantity, $\tau_U$, which is well-defined in these theories. Furthermore, we conjecture that this quantity is larger in the ultraviolet (UV) than in the infrared (IR), i.e. that $\tau_U^{UV}>\tau_U^{IR}$. While we do not prove this inequality in full generality, it is straightforward to show that our conjecture holds in the subset of theories that do not have accidental symmetries. In addition, we subject our inequality to an array of nontrivial tests in theories with accidental symmetries and dramatically different dynamics both in $\CN=1$ and $\CN=2$ supersymmetry and find that our inequality is obeyed. One interesting consequence of this conjecture is that the mixing of accidental symmetries with the IR superconformal $R$ current is bounded by the UV quantity, $\tau_U^{UV}$. To demonstrate the potential utility of this bound, we apply it to the somewhat mysterious $SU(2)$ gauge theory of Intriligator, Seiberg, and Shenker and show that our conjecture, if correct, implies that this theory flows in the IR to an interacting superconformal field theory.

\bigskip
\Date{September 2011}

\newsec{Introduction}
All known renormalization group (RG) flows in four-dimensions can be thought of as interpolations between two sets of conformal fixed points: one set that describes the short-distance ultraviolet (UV) physics and another set that describes the long-distance infrared (IR) physics.\foot{We ignore subtleties associated with Goldstone bosons.} \foot{This statement is rather generally true in two dimensions \refs{\ZamolodchikovGT, \PolchinskiDY} (see also the illuminating discussion in \DorigoniRA; for a holographic perspective, see \NakayamaWX). In four dimensions, however, it is an open question. For some sufficient conditions under which $R$-symmetric RG fixed points are also conformal in four dimensions, see \AntoniadisGN. In fact, by applying the arguments described in \AntoniadisGN, we will be able to immediately conclude that the $R$-symmetric RG flows discussed below that are initiated by turning on a marginally relevant gauge coupling (without also turning on a non-trivial superpotential) are necessarily flows between conformal fixed points. For the remaining examples, duality and various consistency checks strongly suggest that this picture is still correct. Moving to $4-\epsilon$ dimensions, some non-supersymmetric scale invariant but non-conformal theories have been discussed recently in \FortinKS.}

In the deep UV and the deep IR, the description of the physics typically simplifies (although it remains unsolvable in general), and many important properties of the corresponding conformal field theories (CFTs) can be described by the various conserved currents and the simple numerical coefficients associated with the correlation functions of these currents. For example, the $a$ and $c$ central charges of the CFT---which measure the number of degrees of freedom of the theory---can be defined by considering the one point function of the trace of the stress tensor, $\langle T_{\mu}^{\ \mu}\rangle$, in a curved background and computing the coefficients of the Euler density and the square of the Weyl tensor respectively ($a$ and $c$ can also be defined via certain higher-point functions of the stress tensor in the flat space theory). Other interesting numbers include the coefficients of the two-point functions of the internal symmetry currents, $\tau_{ij}$
\eqn\kCFT{
\langle j_{\mu, i}(x) j_{\nu, j}(0)\rangle={\tau_{ij}\over(2\pi)^4}\left(\partial^2\eta_{\mu\nu}-\partial_{\mu}\partial_{\nu}\right){1\over x^4}~.
}
Unitarity implies that $\tau_{ij}$ is a positive definite matrix. This object measures the amount of matter charged under the various global symmetries and the violation of scale invariance when these symmetries are gauged. In fact, as we will see in the supersymmetric (SUSY) theories we discuss below, these different numbers are often related.

One natural question that emerges from this picture is to understand how these various quantities change under the influence of the RG flow. Indeed, by establishing general relations between the UV and IR values of the coefficients of the various current correlation functions, we can hope to gain some simple understanding of the incredibly complicated dynamics that occur along the flows of generic theories and perhaps find criteria for determining their IR phases.

A particularly important and well-studied aspect of this program is Zamolodchikov's proof \ZamolodchikovGT\ that in two dimensional RG flows, the $c$ charge of the UV CFT, $c_{UV}$, is larger than the corresponding quantity at the IR fixed point, i.e. that
\eqn\cthm{
c_{UV}>c_{IR}~.
}
This proof gives some rigor to the intuition that RG flows are irreversible processes that are characterized by a reduction in the number of degrees of freedom of the theory as one integrates out longer and longer wavelength modes.

More recently, proving part of a conjecture due to Cardy \CardyCWA, Komargodski and Schwimmer \KomargodskiVJ\ generalized this idea to four dimensions\foot{See \refs{\MyersTJ\FreedmanGP-\GirardelloBD} for a holographic approach to Cardy's conjecture.} and showed that $a$ in fact decreases along the flow
\eqn\athm{
a_{UV}>a_{IR}~.
}
The result in \athm\ is very general and can be used to determine the IR phases\foot{See \AppelquistHR\ for another quantity that is conjectured to decrease and to be useful in determining IR phases of certain theories.} of certain gauge theories (note, however, that this inequality will not be directly useful in determining the IR phase of the Intriligator, Seiberg, and Shenker (ISS) theory \IntriligatorRX\ we will study below) even though $a_{IR}$ is incalculable in many examples.\foot{As we will review below, SUSY often dramatically improves the situation. Note that we can, however, compute the central charges in some interesting cases even without SUSY. For example, it is straightforward to carry out this computation in the case of real-world QCD, since it is a free theory of quarks and gluons in the UV and a free theory of pions in the IR.}

In this paper, we extend this program by specializing to four dimensional $R$-symmetric SUSY theories and providing evidence for a new RG inequality that is independent of \athm. We will find that this conjectured inequality, if true, gives a particularly strong handle on the IR phases of $R$-symmetric gauge theories.

The starting point for our construction is the multiplet for the dimension three conserved $R$ current, $\CR_{\mu}$, that any $R$-symmetric theory must posses. Provided the theory also has a Ferrara-Zumino (FZ) multiplet \FerraraPZ,\foot{All the theories we will study have such multiplets. The known theories that do not have such multiplets are those with field-independent FI terms \KomargodskiPC\ or sigma models with nontrivial target space topology \KomargodskiRB.} the $\CR_{\mu}$ multiplet obeys the following superspace conservation equation \KomargodskiRB
\eqn\Rcurrcons{
\bar D^{\dot\alpha}\CR_{\alpha\dot\alpha}=\bar D^2D_{\alpha}U~,
}
where $U$ is a well-defined (gauge invariant and local) real multiplet. The $\CR_{\mu}$ multiplet contains as its lowest component the conserved $R$ current and has as its higher components the supercurrent and the stress tensor. When $U=0$, the theory is superconformal and the corresponding $\tilde\CR_{\alpha\dot\alpha}$ satisfies
\eqn\RCFT{
\bar D^{\dot\alpha}\tilde\CR_{\alpha\dot\alpha}=0~.
}
This multiplet contains, as its lowest component, the superconformal $R$ current \FerraraPZ; the higher components of this multiplet contain the traceless stress tensor and supercurrent.

Before proceeding, we should note that \Rcurrcons\ has an ambiguity. Indeed, $U$ is defined modulo chiral plus anti-chiral terms, $X+\bar X$, since $\bar D^2D_{\alpha}(X+\bar X)=0$. As we will describe below, however, this ambiguity will not be important.

In addition to the above mentioned ambiguity, we can also shift $U$ by a real superfield, $J$, with $J|_{\theta\sigma^{\mu}\bar\theta}\equiv -J_{\mu}$ a conserved non-$R$ current (the conservation condition is equivalent to the superspace constraint $D^2J=0$) to find a new conserved $R$ current $\CR'_{\alpha\dot\alpha}=\CR_{\alpha\dot\alpha}+{2\over3}[D_{\alpha}, \bar D_{\dot\alpha}]J$. This ability to shift the $R$ current superfield corresponds to the well-known fact that in $R$-symmetric theories with flavor (i.e., non-$R$) symmetries, the $R$ symmetry is not unique. Under these transformations, the component supercurrent and stress tensor shift by improvement terms.\foot{Note that if we shift $U$ by an operator, $J$, of the particular form $J=D^{\alpha}\chi_{\alpha}+\bar D_{\dot\alpha}\bar\chi^{\dot\alpha}$ for chiral $\chi_{\alpha}$, the component $R$ current itself shifts by an improvement transformation (i.e., the vector component of $J$ is just an improvement term \DumitrescuCA). However, due to unitarity, such terms will not play a role in the superconformal field theory (SCFT) quantities we compute.}

All the theories we will study are defined by considering a UV fixed point and turning on a relevant deformation (possibly in conjunction with a dangerously irrelevant deformation) that preserves a particular $R$ symmetry of the short-distance SCFT (we can also turn on vacuum expectation values for some set of scalar operators that are neutral under this $R$ symmetry).

The $R$ symmetry of the deformed theory descends from a conserved $R$ current superfield of the undeformed UV fixed point, which we write as $\CR_{\mu}^{UV*}$.\foot{The notation \lq $^*$' is to emphasize that the corresponding operator is defined at a conformal fixed point. In this case, the fixed point is just the undeformed UV theory.} We then denote the multiplet related to this superfield via \Rcurrcons\ as $U^{UV*}$. In the UV SCFT, $U^{UV*}|_{\theta\sigma^{\mu}\bar\theta}\equiv -U^{UV*}_{\mu}$ must be conserved (i.e., at the level of superfields, $D^2U^{UV*}=0$), since $\CR_{\mu}^{UV*}$ is related to the conserved superconformal $R$ current multiplet, $\tilde R_{\mu}^{UV}$, by a shift of the type discussed above with $U^{UV*}$ playing the role of $J$.

After turning on the relevant deformation, $\CR_{\mu}^{UV}$ is conserved by construction even though $U^{UV}_{\mu}$ is {\it not}. We can then follow the RG evolution of the non-conserved $U^{UV}$ current superfield using the conserved $R$ current multiplet (many crucial aspects and applications of this RG evolution were discussed recently in \AbelWV, along with an application that relates to previous studies in \LutyQC) and find that as we go to the deep IR, the $U^{UV}$ multiplet flows, modulo chiral plus anti-chiral terms, to a conserved current superfield of the IR fixed point, $U^{IR*}$. The corresponding component conserved current, $U_{\mu}^{IR*}$, measures the difference between the IR superconformal $R$ current, $\tilde R^{IR}_{\mu}$, and the IR limit of our RG-conserved $R$ current, $\CR_{\mu}^{IR}$ (as discussed in footnote 2, we can only prove that $U_{\mu}^{IR*}$ is conserved and that the IR theory is therefore superconformal in a subset of the theories under consideration, but we will assume that this statement holds more generally in our class of theories). More precisely, we have that
\eqn\UUVIR{
U_{\mu}^{UV*, IR*}\equiv -U^{UV*, IR*}|_{\theta\sigma^{\mu}\bar\theta}={3\over2}\left(\CR_{\mu}^{UV*, IR*}-\tilde \CR^{UV, IR}_{\mu}\right)~.
}

This discussion suggests a simple and natural numerical quantity to compute at the conformal endpoints of $R$-symmetric RG flows: the coefficients of the two-point functions of the conserved $U^{UV*, IR*}$ multiplets.\foot{Coefficients of other natural two-point functions in this context correspond to well-studied quantities. For example, the coefficient of the superconformal $R$ current two-point function corresponds to $c$.} However, these quantities are not well-defined in general without further input. Indeed, if the relevant deformations we turn on in the UV preserve some flavor symmetries of the UV SCFT, then, as discussed above, we have an infinite family of $R$-symmetries and corresponding currents.

One particularly natural choice for the pairing $(\CR_{\mu}^{UV}, U^{UV})$ is the one determined by performing $a$-maximization \IntriligatorJJ\ in the deformed UV theory (note that here we are not using $a$-maximization at the fixed points\foot{This is slightly reminiscent of the generalization in \KutasovUX.}).
Indeed, modulo some subtleties and exceptions we will deal with in the next section, this procedure yields an RG-conserved $R$ current (and a corresponding $U$ operator) that descends from an $R$ current (and $U$ operator) of the undeformed UV fixed point. We denote these UV SCFT operators as $(\CR_{\mu, \rm vis}^{UV}, U_{\rm vis}^{UV})$.\foot{The subscript \lq vis' is to remind us that these operators are defined in terms of the visible (short distance) degrees of freedom and that $\CR_{\rm vis}$ is a symmetry of the full theory.} We drop the superscript \lq$ ^*$' when writing these operators, because it is understood that they are defined at the undeformed UV fixed point. We write the two-point functions of the corresponding UV and IR $U_{\mu, \rm vis}$ operators as follows
\eqn\UUdefn{
\langle U_{\mu, \rm vis}^{UV, IR}(x) U_{\nu, \rm vis}^{UV, IR}(0)\rangle={\tau_U^{UV, IR}\over(2\pi)^4}\left(\partial^2\eta_{\mu\nu}-\partial_{\mu}\partial_{\nu}\right){1\over x^4}~.
}
Let us emphasize that these two-point functions are to be interpreted as being evaluated in the (undeformed) UV and IR SCFTs respectively.

In many theories, the coefficient $\tau_U$ is smaller in the IR than in the UV.\foot{As discussed above, in writing \UUVIR\ and \UUdefn\ we have dropped possible chiral plus anti-chiral operators. Such operators (including goldstons bosons) only contribute to the $\partial_{\mu}\partial_{\nu}$ term in the full $\langle U_{\mu, \rm vis}^{UV, IR}(x)U^{UV, IR}_{\nu, \rm vis}(0)\rangle$ correlator. Therefore, we can alternatively define $\tau_{U}$ to be the coefficient of the $\eta_{\mu\nu}\partial^2x^{-4}$ part of the full correlator.} For example, in any perturbative theory or, more generally, in any theory without accidental symmetries in the IR\foot{SQCD in the conformal window, $3N_c/2<N_f<3N_c$, is widely believed to be an example of such a theory \SeibergPQ.} 
\eqn\perttau{
\tau_U^{UV}>0=\tau_U^{IR}~.
}
The fact that $\tau_U>0$ follows from unitarity and the fact that $U^{UV}_{\mu, \rm vis}\ne0$ (since $R_{\mu, \rm vis}^{UV}\ne\tilde R_{\mu}^{UV}$). On the other hand, $\tau_U^{IR}=0$ in this class of theories since the $R$ current defined by $a$-maximization in the deformed UV theory, $\CR^{UV}_{\mu, \rm vis}$, flows to the superconformal $R$ current in the deep IR and so \UUVIR\ guarantees that $U^{IR}_{\rm vis}=0$ (see also the discussion in section six of \AbelWV).

More generally, however, the IR phases of $R$-symmetric gauge theories may have accidental symmetries. If these symmetries mix with the superconformal $R$ current, then the current defined by $a$-maximization, $\CR_{\mu, \rm vis}^{UV}$, does not flow to the superconformal $R$ current, $\tilde\CR_{\mu}^{IR}$, and $U_{\rm vis}^{UV}$ does not flow to zero. In fact, $U^{UV}_{\rm vis}$ flows to an accidentally conserved current superfield of the IR SCFT,\foot{Strictly speaking, $U^{UV}_{\rm vis}$ can flow to a linear combination that also includes a component corresponding to a conserved current superfield of the full theory. However, this subtlety will not affect our discussion in any way.} i.e., $U_{\rm vis}^{UV}\to J_{SCFT}^{IR}$ where
\eqn\UvisIRgen{
D^2J_{SCFT}^{IR}=0~.
}
In this case $\tau_U^{IR}\ne0$, and  it is no longer obvious that $\tau_U^{UV}>\tau_U^{IR}$. Indeed, this inequality no longer follows from $a$-maximization, and it is not implied by the $a$-theorem \athm.

Surprisingly, we will find strong evidence that even in theories with accidental symmetries
\eqn\kconjpaper{
\tau_U^{UV}>\tau_U^{IR}~.
}
Since $\tau_U^{IR}$ measures the mixing of the IR superconformal $R$ current with the accidental symmetries,\foot{One way to make this notion precise is to note that minimization of the coefficient of the $R$ current two point function \BarnesBM\ implies that $16c_{IR}/3=\tau^{IR}_{RR}=\tau_{\rm vis}^{IR}-4\tau_U^{IR}/9$, where $\tau^{IR}_{RR}$ is the coefficient of the IR superconformal $R$ current two-point function, and $\tau^{IR}_{\rm vis}$ is the coefficient of the two-point function of the IR limit of $\CR_{\mu, \rm vis}$. Therefore, the two point function of the IR superconformal $R$ current receives two contributions: a contribution from a visible symmetry (i.e., a symmetry of the full theory) and a contribution from an accidental symmetry that appears in the long-distance theory.} it follows that \kconjpaper\ constitutes a (conjectured) bound on this mixing in a large class of $R$-symmetric theories.

More precisely, this conjecture applies to the conformal endpoints of all $R$-symmetric theories that have an FZ multiplet and that are defined by turning on relevant deformations of a UV fixed point (along with possible dangerously irrelevant deformations) and / or turning on a set of $R$-symmetry preserving vevs.

We will not prove \kconjpaper, but we will test it under a variety of deformations in theories with very different dynamics. At a heuristic level, our conjecture seems plausible since the smaller $\tau_U^{UV}$ is, the more approximately conformal we expect the deformed UV theory to be, and the more likely that the IR phase of the theory is an SCFT with no confined fields and no accidental symmetries.\foot{See \IntriligatorIF\ for a different conjectured criterion to determine whether a theory is IR conformal or confining.} Of course, it would be interesting to make these ideas more precise. In this paper, we merely hope to collect some facts surrounding these claims, and we postpone an attempt to prove our conjecture for the future (provided, of course, that there isn't a counterexample).

Before proceeding, we should emphasize that while  $\tau_U^{UV}$ is a quantity in the UV SCFT, it is not intrinsically defined in the UV SCFT (unlike the $a$ and $c$ charges). Rather, it is defined once we have in mind a particular relevant deformation (along with a possible dangerously irrelevant one) and / or a scalar vev of the SCFT. As a result, it cannot count the number of degrees of freedom in the deep UV (after all, it is just the coefficient of a particular conserved flavor current two-point function of the UV SCFT), and therefore even if we managed to prove that  $\tau_U^{UV}>\tau_U^{IR}$, we would not be able to deduce anything about the irreversibility of the RG flow. Note, however, that $\tau_U^{UV}$ is independent of the details of the RG flow since it does not depend on the precise coefficients of the deformations or the particular value of the scalar vevs that start the flow.

The plan for the rest of this paper is as follows. In the next section, we introduce the technical details of our conjecture. After laying this groundwork, we thoroughly test our ideas in the context of $\CN=1$ $SU(N_c)$ SQCD. We consider examples of SQCD flows in which both endpoints of the flow are free, one endpoint is free, and both endpoints are interacting (i.e., for flows between interacting IR fixed points in the conformal window). We then proceed to consider our conjecture in the case of a particularly illustrative s-confining example. In order to make the paper more digestible, we consign the tests of our conjecture in a plethora of additional theories ($SO(N_c)$ SQCD, $Sp(N_c)$ SQCD, $\CN=2$ Super Yang-Mills (SYM), various interacting SCFTs  with accidental symmetries, additional s-confining examples, and some more complicated free magnetic theories) to the various appendices. Finally, we apply our conjecture to the mysterious ISS theory \IntriligatorRX\ and find that its IR phase must be interacting conformal (if our conjecture is correct). We conclude with some thoughts about future directions.
 
\newsec{Computing $\tau_U$}
In this section, we would like to describe the procedure for computing $\tau_U$ in greater detail before proceeding to study SQCD in the next section. We conclude this section with the particularly simple but nonetheless informative example of the free massive chiral superfield.

As described in the introduction, we start our analysis at the UV fixed point, where we can use $a$-maximization to compute the superconformal $R$ current \IntriligatorJJ. This computation proceeds as follows: we consider the full set of non-$R$ symmetry currents of the UV SCFT, $\left\{J_{\mu,i}^{UV*}\right\}$, and we pick a particular reference $R$ current, $\CR^{(0)*}_{\mu, UV}$. Then, we define a trial $R$ symmetry, $\CR^{t*}_{\mu, UV}$
\eqn\RtUV{
\CR^{t*}_{\mu, UV}=\CR^{(0)*}_{\mu, UV}+\sum_it^iJ_{\mu, i}^{UV*}~,
}
where the $t^i$ are real numbers. The superconformal $R$ symmetry, $\tilde \CR_{\mu}^{UV}$, is defined by the unique set of $t^i=t^i_*$ such that the corresponding trial $\tilde a^t$ function\foot{Recall that in SCFTs, we can compute $a$ in terms of the anomalies of the superconformal $R$ current, $\tilde R_{\mu}$ \refs{\AnselmiAM, \AnselmiYS}. Here we define $a={3\over32}\tilde a={3\over32}\left(3\Tr \tilde \CR^3-\Tr \tilde \CR\right)$. The charges appearing in the trace are understood to be over massless $R$-charged fermions.}
\eqn\trialaUV{
\tilde a_{UV}^t=3\Tr \left(\CR^{t*}_{UV}\right)^3-\Tr \CR^{t*}_{UV}~,
}
is maximised, i.e.
\eqn\amaxUV{
\partial_{t^i}{\tilde a_{UV}}^t|_{t^i=t^i_*}=0, \ \ \ \partial^2_{t^it^j}{\tilde a_{UV}}^t|_{t^{i,j}=t^{i,j}_*}<0~.
}

We then imagine deforming the UV theory by adding a set of relevant operators (possibly along with a set of dangerously irrelevant operators) and / or turning on a set of scalar vevs that preserves an $R$ symmetry to the Lagrangian. As described above, the resulting $R$ symmetry is not unique in general since the deformation may respect a nontrivial subset of the flavor symmetries of the UV SCFT. In such a case, there is a non-vanishing subset of flavor currents of the UV SCFT $\left\{\hat J_{\mu, a}^{UV*}\right\}\subset\left\{J_{\mu, i}^{UV*}\right\}$ that gives rise to conserved currents of the full RG flow. Therefore, the most general trial $R$ current for the full theory is
\eqn\RRG{
\CR^{t, UV}_{\mu}=\CR^{(0), UV}_{\mu}+\sum_a{\hat t}^a\hat J^{UV}_{\mu, a}~,
}
where the conserved $\hat J_{\mu, a}^{UV}$ currents descend from the corresponding $\hat J_{\mu, a}^{UV*}$ currents of the UV SCFT. Maximizing $\tilde a^t$ over the restricted set of currents compatible with the deformation (and any non-zero vevs) yields a set of coefficients $\hat t^a=\hat t^a_*$ that defines
\eqn\Rvis{
\CR_{\mu}^{UV}=\CR^{(0), UV}_{\mu}+\sum_a{\hat t}^a_*\hat J^{UV}_{\mu, a}~.
}
This procedure determines the corresponding $U^{UV}$ multiplet via \Rcurrcons.

The $(\CR_{\mu}^{UV}, U^{UV})$ operators we have described above, descend from the operators $(\CR_{\mu, \rm vis}^{UV}, U_{\rm vis}^{UV})$ of the undeformed UV SCFT. It is these latter operators that we will study below.

Before proceeding, however, we should note that there is one caveat to our above discussion. Indeed, it may happen that $a$-maximization in the deformed UV theory does not determine the mixing of $\CR^{UV}_{\mu, \rm vis}$ with a nontrivial subset of the conserved flavor currents of the deformed theory, $\left\{\tilde J_{\mu, A}^{UV}\right\}\subset\left\{\hat J_{\mu, a}^{UV}\right\}$. In the absence of accidental symmetries, the physical interpretation of this statement is that the $\left\{\tilde J_{\mu, A}^{UV}\right\}$ currents flow to zero in the deep IR (one way this can happen is if the corresponding symmetries only act on massive particles). In other situations, it may happen that $a$-maximization gives rise to complex $\hat t^a$ and so the IR theory must contain accidental symmetries that mix with the IR superconformal $R$ current. In any of these cases, we fix the ambiguity in the corresponding $(\CR_{\mu, \rm vis}^{UV}, U_{\mu, \rm vis}^{UV})$ pair by demanding that 
\eqn\seconddem{
\langle U^{UV}_{\mu, \rm vis}(x)\tilde J^{UV*}_{\nu, A}(0)\rangle=0~,
}
for all $A$. Note that this modification does not alter our proof in the introduction that $\tau_U^{UV}>\tau_U^{IR}$ in the subset of theories without accidental symmetries.

In general, it is rather difficult to obtain explicit expressions for $\CR_{\mu, \rm vis}^{UV}$, $U_{\rm vis}^{UV}$, and $\tilde \CR_{\mu}^{UV}$. Of course, the situation simplifies dramatically if the theory we want to study is asymptotically free (we will discuss the case of an interacting UV SCFT shortly). In that case, $\tilde \CR_{\mu}^{UV}$ assigns $R$-charges $2/3$ to all the chiral superfields of the theory, we can solve for $\CR_{\mu, \rm vis}^{UV}$ in terms of some free chiral superfields, $\Phi_i$, and we find the following simple expression for the $U_{\rm vis}^{UV}$ charges from \UUVIR\
\eqn\UUVafree{
U^{UV}_{\rm vis}(\Phi_i)={3\over2}\left(\CR^{UV}_{\rm vis}(\Phi_i)-{2\over3}\right)~.
}
Our superfield expression for the corresponding multiplet is then $U^{UV}_{\rm vis}=-\sum_iU_{\rm vis}^{UV}(\Phi_i)\Phi_i\bar\Phi_i$. Computing $\tau_U^{UV}$ then reduces to computing the sum of the squares of the charges of $U_{\mu, \rm vis}^{UV}$, i.e.
\eqn\tauUVfree{
\tau_U^{UV}=\Tr\ \left(U^{UV}_{\rm vis}\right)^2~.
}

Next, we study the theory as it flows into the deep IR. If no accidental symmetries emerge along the flow (or, at least, if no accidental symmetries mix with the superconformal $R$ current) $\tau_U^{IR}=0$, and our conjecture is satisfied.

Of course, as discussed in the introduction, $\CR_{\mu, \rm vis}$ generally flows to a conserved $R$ current of the IR SCFT that differs from the superconformal one by an accidentally conserved non-$R$ current.

These accidental symmetries are of two general types \KutasovIY\ (see also the interesting discussion in \LeighDS): the first type is detectable in the deformed UV theory and the second type is not. For example, detectable accidental symmetries appear in theories in which some of the composite chiral gauge invariant operators built out of the UV fields, $\CO_I$, have $\CR_{\rm vis}(\CO_I)<2/3$.\foot{The canonical example of this situation is the meson field, $M$, in SQCD with $N_f<3N_c/2$.} Clearly, in such a case, $\CR_{\mu, \rm vis}^{IR}$ cannot correspond to the IR superconformal $R$ current since then $d(\CO_I)<1$ and the unitarity bound would be violated.\foot{As suggested in \PolandWG, one could also try to incorporate bounds of the types found in \refs{\PolandWG, \VichiUX}.} In general, it is believed (although it has never been proven) that the correct description of the physics in this situation is that  the $\CO_I$ decouple from the IR theory and become free fields with superconformal $R$-charge $2/3$, and so the accidental symmetries under which each $\CO_I$ separately transforms mix with the IR superconformal $R$ current. The second type of accidental symmetry is even more problematic because there is no obvious way to see that it is present without a full description of the IR physics.\foot{An example of this situation is $N_f=N_c+1>3$ SQCD. In this theory, the baryons have $\CR_{\rm vis}(B)=\CR_{\rm vis}(\tilde B)>2/3$ but are nonetheless free in the deep IR.}

For detectable accidental symmetries of the type we have described above, one can modify $\tilde a^t$ to take into account the (assumed) decoupling of the $\CO_I$ operators \KutasovIY
\eqn\auvdef{\eqalign{
\tilde a^t&\to \tilde a^t-\sum_I\dim(\CO_I)\left[\left(3\left(\CR^t(\CO_I)\right)^3-\CR^t(\CO_I)\right)-\left(3\left(-{1\over3}\right)^3-\left(-{1\over3}\right)\right)\right]\cr&=\tilde a^t+{1\over9}\sum_I\dim(\CO_I)\left(2-3\CR^t(\CO_I)\right)^2\left(5-3\CR^t(\CO_I)\right)~,
}}
where ${\rm dim}(\CO_I)$ is the number of degrees of freedom in $\CO_I$ (not the scaling dimension). The heuristic idea behind \auvdef\ is that one should replace the $\CR^t$ dependence of the contributions from the $\CO_I$ operators with the contributions from free fields. Maximizing this $\tilde a^t$ defines a new central charge, $\hat a$. Assuming that there are no additional accidental symmetries, $\hat a =a_{IR}$.

Note, however, that we will not use the deformed trial $a$ function in \auvdef\ to define $(\CR_{\mu, \rm vis}^{UV}, U_{\rm vis}^{UV})$. There are two reasons for this fact. First, we would like to treat all types of accidental symmetries as uniformly as possible. Second, adding the deformation in \auvdef\ sometimes prevents one from constructing an $\CR_{\mu, \rm vis}$ current using $a$-maximization. The reason for this is simple. It may happen that the IR consists exclusively of decoupled chiral gauge invariant operators that violate the unitarity bound. In that case, \auvdef\ simply (and correctly) reproduces the central charge for the free theory of the confined $\CO_I$ operators. In particular, the dependence of $\tilde a^t$ on the various $t^i$ coefficients parameterizing $R$ current mixing with the flavor symmetries of the UV theory drops out since the deformation in \auvdef\ replaces the $\CR^t$-dependent $\CO_I$ contributions with contributions from free fields of $R$-charge $2/3$. We will see that such a situation arises in the $SU(7)$ s-confining theory of section 4.\foot{In fact, it also arises in $N_f=N_c$ SQCD.}

Note, however, that if the IR theory contains additional, interacting IR fields, then \auvdef\ will give rise to an RG-conserved $R$ current, $\hat \CR_{\mu}^{UV},$ that will flow in the IR to a current that coincides with the IR superconformal $R$ current in the interacting sector (although it differs from the free superconformal $R$ current in the decoupled sector), i.e., we have the flow $\hat \CR_{\mu}^{UV}\to\tilde \CR_{\mu}^{IR}|_{\rm int}+\cdot\cdot\cdot$, where $\tilde \CR_{\mu}^{IR}|_{\rm int}$  is the restriction of the IR superconformal $R$ current to the interacting sector, and the ellipses include contributions from the decoupled, free sector. Using the fact that SUSY guarantees $\tau_U=-3\Tr \left(\tilde \CR UU\right)$ for conserved currents in SCFTs, we can compute $\tau_U^{IR}$
\eqn\kUIRintfree{
\tau_U^{IR}=-{27\over4}\Tr\ \hat \CR^{UV}(\CR_{\rm vis}^{UV}-\hat \CR^{UV})^2+{27\over4}\Tr|_{\rm free}\ \hat \CR^{IR}(\CR^{IR}_{\rm vis}-\hat \CR^{IR})^2+\Tr|_{\rm free}\left(U^{IR}_{\rm vis}\right)^2~.
}
In this formula, the first trace is understood, via anomaly matching, as being over the free UV fermions, and the remaining two traces are over the fermions of the decoupled (free) IR sector (the first two terms effectively capture the contribution of the interacting IR sector). Of course, in theories that are completely free in the IR, like SQCD in the free magnetic phase, \kUIRintfree\ simplifies and we find only a contribution from the last trace in \kUIRintfree
\eqn\tauIRfree{
\tau_U^{IR}=\Tr\left(U^{IR}_{\rm vis}\right)^2~.
}

In our discussion thus far, we have assumed that the UV theory is free. In fact, we can still make some concrete statements when the RG flow is between certain (strongly) interacting fixed points. Indeed, it may happen that both the UV and IR theories are RG descendants of some parent free theory (as happens in flows between interacting fixed points in the conformal window of SQCD)---for simplicity we will assume that all the symmetries of the interacting UV fixed point are visible in this free parent theory. Then, proceeding as before, we can turn on a deformation of the interacting UV fixed point and define an $\CR_{\mu, \rm vis}$ current for the flow to the IR theory. Using 't Hooft anomaly matching, we can then compute
\eqn\kanom{
\tau_U^{UV}=-3\Tr\ \tilde \CR^{UV}_{p} U^{UV}_{\rm vis, p}U^{UV}_{\rm vis, p}~,
}
where the trace is understood as being evaluated for fermions in the parent (free) theory (i.e., we use 't Hooft anomaly matching since the UV superconformal $R$ current and the $\CR_{\rm vis}$ current for the flow to the IR can both be thought of as IR limits of $R$-symmetries of the free parent theory). In the IR we then find, in analogy with \kUIRintfree
\eqn\kUIRintfreecomp{
\tau_U^{IR}=-{27\over4}\Tr\ \hat R^{UV}_{\rm p}(\CR_{\rm vis, p}^{UV}-\hat R^{UV}_{\rm p})^2+{27\over4}\Tr|_{\rm free}\ \hat R^{IR}(\CR^{IR}_{\rm vis}-\hat R^{IR})^2+\Tr|_{\rm free}\left(U^{IR}_{\rm vis}\right)^2~,
}
where here the first trace is to be evaluated for fermions in the parent (free) theory.

In the next subsection we put some of the simplest aspects of the above machinery to work, and then in the next section we move on to a study of SQCD.
\subsec{Free massive chiral multiplet}
As a simple check of some of the above ideas, consider a free theory of a single chiral multiplet, $\Phi$. Let us deform the theory by adding a mass
\eqn\Wmass{
W=m\Phi^2~.
}
The theory then flows to a trivial theory in the IR. The deformation in \Wmass\ fixes $\CR_{\rm vis}^{UV}(\Phi)=1$ and $U_{\rm vis}^{UV}(\Phi)=1/2$. As a result $\tau_U^{UV}={1\over4}$, and $\tau_U^{IR}=0$ since the IR theory is trivial. Therefore,
\eqn\kconjphi{
\tau_U^{UV}>\tau_U^{IR}~,
}
and so our conjecture holds in this simple case.\foot{Note that we can also consider the theory of a free pair of chiral superfields, $\Phi_{1,2}$, and imagine deforming it by turning on $W=m\Phi_1\Phi_2$. In this case, the RG-flow preserves the symmetry under which $\Phi_1$ and $\Phi_2$ have opposite charges. As discussed in the previous subsection, $a$-maximization does not determine the mixing of $\CR_{\rm vis}^{UV}$ with this symmetry since the corresponding current, $J_{\mu}^{UV*}$, flows to zero at the (trivial) IR fixed point. Imposing $\langle U_{\mu, \rm vis}^{UV}(x)J_{\nu}^{UV*}(0)\rangle=0$ as in \seconddem, however, fixes $U_{\rm vis}^{UV}(\Phi_{1,2})=1/2$ and $\tau_U^{UV}=1/2>0=\tau_U^{IR}$.} We now move on to SQCD.

\newsec{SQCD}
We begin by analyzing our conjecture in the context of $SU(N_c)$ $\CN=1$ SQCD with $N_f<3N_c$. Note that our procedure for constructing $\CR_{\rm vis}$ treats all the $Q$ and $\tilde Q$ fields symmetrically, and so $\CR_{\rm vis}^{UV}(Q)=\CR_{\rm vis}^{UV}(\tilde Q)=1-{N_c\over N_f}$. We first consider flows starting from the free UV fixed point, and therefore \UUVafree\ tells us that  $U_{\rm vis}^{UV}(Q)=U_{\rm vis}^{UV}(\tilde Q)={1\over2}-{3N_c\over 2N_f}$.

Now, recalling that the theories with $N_f<N_c$ do not have stable vacua at finite values in field space, we begin by analyzing the case $N_f=N_c$ and work our way up in $N_f$.

\subsec{Deformed moduli space: $N_f=N_c$}
In this case, the theory confines with chiral symmetry breaking and $U_{\rm vis}^{UV}(Q)=U_{\rm vis}^{UV}(\tilde Q)=-1$. Our formula in \tauUVfree\ tells us that
\eqn\kUVdeformed{
\tau_U^{UV}=2N_c^2~.
}
On the other hand, the IR consists of $N_c^2$ mesons, $M$, and two baryons $B$ and $\tilde B$ with $\CR_{\rm vis}^{IR}(M)=\CR_{\rm vis}^{IR}(B)=\CR_{\rm vis}^{IR}(\tilde B)=0$ and $U_{\rm vis}^{IR}(M)=U_{\rm vis}^{IR}(B)=U_{\rm vis}^{IR}(\tilde B)=-1$ satisfying 
\eqn\defmodspace{
\det M+B\tilde B=\Lambda^{2N_c}~.
}
Therefore, the number of massless degrees of freedom in the IR is always less than $N_c^2+2$ since the deformed moduli space constraint removes degrees of freedom. We then find from \tauIRfree\ that
\eqn\kIRdeformed{
\tau_U^{IR}<N_c^2+2~.
}

Note that in this example, $U^{IR}$ contains holomorphic plus anti-holomorphic terms built out of the Goldstone modes arising from the deformed moduli space constraint in \defmodspace\ \AbelWV. Therefore, in computing $\tau_U^{IR}$, we should either drop these terms or recall the discussion in footnote 12 and compute the coefficient of $\eta_{\mu\nu}\partial^2x^{-4}$ in $\langle U_{\mu}^{IR}(x)U_{\nu}^{IR}(0)\rangle$.

We conclude by noting that this result is compatible with our conjecture since
\eqn\inneqdef{
\tau^{UV}_U>\tau^{IR}_U~.
}

\subsec{$s$-confinement: $N_f=N_c+1$}
In this case, the theory at the origin of the moduli space confines without chiral symmetry breaking. Since $U(Q)=U(\tilde Q)=(1-2N_c)/(2+2N_c)$, we see that
\eqn\kUVsconfn{
\tau_U^{UV}={N_c(1-2N_c)^2\over 2(1+N_c)}~.
}
The IR consists of $(N_c+1)^2$ mesons, $M$, and $2(N_c+1)$ baryons $B$ and $\tilde B$. These operators have $\CR_{\rm vis}(M)={1-2N_c\over 1+N_c}, \ \ \ \CR_{\rm vis}(B)=\CR_{\rm vis}(\tilde B)={N_c\over2}{1-2N_c\over 1+N_c}$ and $U(M)=-1+{3\over N_c+1}, \ \ \ U(B)=U(\tilde B)={N_c-2\over 2(N_c+1)}$. Therefore, we find that
\eqn\kIRsconfin{
\tau_U^{IR}={(N_c-2)^2(3+2N_c)\over2(1+N_c)}~.
}
It is easy to see that since $N_c\ge2$ our conjecture holds and
\eqn\kconmpsconf{
\tau_U^{UV}>\tau_U^{IR}~.
}

As an aside, note that the UV coefficient of the two point function of the baryon current is $\tau_B^{UV}=2N_c(N_c+1)$ (where we have taken the UV quarks to have charges $J_B(Q)=-J_B(\tilde Q)=1$). On the other hand, $\tau_B^{IR}=2(N_c+1)N_c^2$. Therefore, $\tau_B^{UV}<\tau_B^{IR}$, and so the baryon two point function coefficient is not a decreasing quantity in this phase.

In fact, the change in $\tau_B$ doesn't have a definite sign. Indeed, had we simply given large masses to all the UV squarks, we would find a trivial theory in the IR and so we would have that $\tau_B^{UV}>\tau_B^{IR}$ in this case.

\subsec{The free magnetic phase: $N_c+1<N_f\le 3N_c/2$}
The $s$-confining description breaks down for $N_f=N_c+2$. This is just as well because $\tau_U^{UV}={2N_c(N_c-1)^2\over N_c+2}$ in this case, while $\tau_U$ evaluated over the $(N_c+2)^2$ mesons and $(N_c+2)(N_c+1)$ baryons yields $\tau_U^{\rm conf}={5N_c^3-10N_c^2-4N_c+36\over N_c+2}$. It is easy to see that in this case $\tau_U^{UV}<\tau_U^{\rm conf}$ for all $N_c\ge2$, which would violate our conjecture.

Fortunately, the Seiberg dual variables yield the correct IR description, and provide additional confirmation for our conjecture. To see this, note that 
\eqn\tauUVfm{
\tau_U^{UV}={N_c(N_f-3N_c)^2\over 2N_f}~,
}
while the free magnetic description has fields with $\CR_{\rm vis}$-charges $\CR_{\rm vis}^{IR}(M)=2\left(1-{N_c\over N_f}\right),  \ \CR_{\rm vis}^{IR}(q)=\CR^{IR}_{\rm vis}(\tilde q)={N_c\over N_f}$ and $U$-charges $U^{IR}_{\rm vis}(M)=2-{3N_c\over N_f}, \ U^{IR}_{\rm vis}(q)=U^{IR}_{\rm vis}(\tilde q)=-1+{3N_c\over2N_f}$. Using this data, we find the IR two-point function coefficient
\eqn\kIRfm{
\tau_U^{IR}={(3N_f-N_c)(3N_c-2N_f)^2\over 2N_f}~.
}
For $N_f$ in the free magnetic range it is then easy to verify that, as desired
\eqn\deltakfm{
\tau_U^{UV}>\tau_U^{IR}~.
}
In fact, this  inequality holds until $N_f\sim1.79 N_c$, which is within the conformal window (where the above description breaks down). Unfortunately, unlike in the transition between confining and free magnetic variables, our inequality doesn't predict the precise onset of the conformal window (although it isn't too far off).

\subsec{The conformal window: $3N_c/2<N_f<3N_c$}
Our conjecture is trivially true in the conformal window. Indeed, this follows from the discussion in the introduction since there are (believed to be) no accidental symmetries in the IR and so $\CR_{\rm vis}$ flows to the IR superconformal $R$ current and hence
\eqn\kconfwin{
\tau_U^{UV}>0=\tau_U^{IR}~.
}
This agreement is, in some sense, much less impressive than the agreement in the confining and free magnetic phases discussed above. However, we will subject our conjecture to much more complicated tests in the conformal window in the next two subsections.

\subsec{Relevant deformations in the conformal window}
In this subsection we consider deforming the SQCD fixed points by adding relevant deformations. It follows from the discussion in the above subsections (including  the discussions in footnote 22 and around \seconddem) that adding mass terms to the asymptotically free theory and turning on the gauge coupling results in flows that satisfy $\tau_U^{UV}>\tau_U^{IR}$.

A much more nontrivial check of our conjecture is to imagine starting from an interacting fixed point in the conformal window and deforming the theory by adding
\eqn\WSQCDmass{
W=\lambda Q_a\tilde Q^a~,
}
where $a=1,...,k$. Note that $\tilde \CR(Q_a\tilde Q^a)<2$, and so \WSQCDmass\ is a relevant deformation of the interacting fixed point. If $k$ is sufficiently small (i.e., if $k< N_f-{3\over2}N_c$), then the theory flows to another fixed point in the conformal window. It follows from our above discussion that $\tau_U^{UV}>\tau_U^{IR}$ in this case (we must recall \seconddem\ to define $\tau_U^{UV}$ for all $k>0$). If $k=N_f-{3\over2}N_c$, we still have $\tau_U^{UV}>\tau_U^{IR}$ because $\CR_{\rm vis}$ flows to the free superconformal $R$ current.

Next, consider the case $N_f-N_c-1>k>N_f-{3\over2}N_c$. At the interacting UV fixed point, we use \kanom\ and the discussion around \seconddem\ to find
\eqn\tauSCFT{
\tau_U^{UV}={27k N_c^4\over 2(N_f-k)N_f^2}~.
}
On the other hand, at the free magnetic fixed point with $N_f-k$ flavors, we use \kIRfm\ and find
\eqn\tauIRmod{
\tau_U^{IR}={(3(N_f-k)-N_c)(3N_c-2(N_f-k))^2\over2(N_f-k)}~.
}
It is then straightforward to verify that
\eqn\taumassfm{
\tau_U^{UV}>\tau_U^{IR}~.
}

Let us now discuss the case $k=N_f-N_c-1$. At the interacting UV fixed point, the expression in \tauSCFT\ still applies. At the IR fixed point, we use \kIRsconfin\ and find that
\eqn\taumasssconf{
\tau_U^{UV}>\tau_U^{IR}~.
}

Finally, consider the case $k=N_f-N_c$. At the interacting UV fixed point, we again use \tauSCFT, while at the IR fixed point we use \kIRdeformed. It is straightforward to see that
\eqn\taumassconf{
\tau_U^{UV}>\tau_U^{IR}~,
}
as desired.

\subsec{Higgsing}
Another nontrivial check of our conjecture is to verify that it is compatible with RG flows involving Higgsing. From the discussion above, it is simple to check that SQCD RG flows starting from the (partially) Higgsed asymptotically free fixed points are compatible with our conjecture.

Indeed, for concreteness, suppose that the first $k$ flavors, $Q_a^i, \tilde Q^a_i$ ($a=1,..., k$ and $i=1,...,N_c$), acquire (distinct) vevs with, say, $\langle Q_a^a\rangle=\langle \tilde Q_a^a\rangle=v_a$. Then, the theory breaks up into an $SU(N_c-k)$ SQCD theory with $N_f-k$ flavors, $\CQ_A^i$, $\tilde{\CQ}_i^A$, and $k^2$ singlets, $S_I$ ($I=1,...,k^2$), under all the symmetries of the reduced SQCD sector along with $k$ gauge singlets, $\Phi_a$, transforming in the ${\bf N_f-k}$ of $SU(N_f-k)_L$ and $k$ gauge singlets, $\tilde\Phi^a$, transforming in the ${\bf \bar{N_f-k}}$ of $SU(N_f-k)_R$. We can consruct an $\CR_{\rm vis}$-symmetry by demanding that it leave the vacuum invariant. Doing so, we find $\CR_{\rm vis}(\CQ)=\CR_{\rm vis}(\tilde\CQ)=\CR_{\rm vis}(\Phi)=\CR_{\rm vis}(\tilde\Phi)=1-{N_c-k\over N_f-k}, \  \CR_{\rm vis}(S)=0$ (note that for the case $k=N_c$ we use \seconddem).

In the deep IR, we expect to find a decoupled theory involving the singlets and a sector that describes the IR of $SU(N_c-k)$ SQCD with $N_f-k$ flavors. Since the singlet sector is free in the UV and the IR, it follows that its contribution to $\tau_U$ is the same both in the IR and the UV. Furthermore, we have already shown that $\tau_U$ decreases in the reduced SQCD sector, and so $\tau_U^{UV}>\tau_U^{IR}$ for the full theory.

A more interesting case occurs when we imagine starting from a fixed point in the conformal window. Provided that $k<{\rm min}\left((3N_c-N_f)/2, N_c-1\right)$, the theory flows to a new (more weakly coupled) fixed point in the conformal window. Using \kanom\ we find
\eqn\tauUVhiggsconf{
\tau_U^{UV}={27kN_c^2(N_c-N_f)^2\over2(N_f-k)N_f^2}~.
}
On the other hand, in the deep IR, we find
\eqn\tauIRhiggsconf{
\tau_U^{IR}={k(2k^2+N_f^2(1-3N_c/N_f)^2+6k N_f(1-2N_c/N_f))\over2(N_f-k)}~.
}
It is easy to check that these results are consistent with our conjecture
\eqn\tauconj{
\tau_U^{UV}>\tau_U^{IR}~.
}

If $(3N_c-N_f)/2\le k\le N_c$, then the end point of the flow is an IR free theory. The two point function coefficient in the UV is again as in \tauUVhiggsconf, but now in the IR we find
\eqn\tauIRfreehiggsconf{
\tau_U^{IR}=-{2k^3-N_f^3(1-3N_c/N_f)^2(N_c/N_f)+4kN_fN_c(3N_c/N_f-1)-2k^2N_f(1+2N_c/N_f)\over2(N_f-k) }~.
}
It is again straightforward to check that
\eqn\tauconjconffreeSQCD{
\tau_U^{UV}>\tau_U^{IR}~.
}

In the next section we will check our results in a more intricate s-confining theory and illustrate several additional points described in the introductory sections.

\newsec{An illustrative $s$-confining example}
In the previous section, we studied our conjecture in the case of $SU(N_c)$ $\CN=1$ SQCD and found that it held in the presence of accidental symmetries as well as in the presence of the various different types of deformations we turned on. While we believe these results to be highly nontrivial, the role of $a$-maximization was somewhat obscured by the $Q\leftrightarrow\tilde Q$ interchange symmetry of the theory. Therefore, in this section we consider a more complicated $s$-confining theory and use it to illustrate various salient points regarding the relationship between $a$-maximization and our conjecture. We also use this example to clarify the role of irrelevant deformations in our construction.

To that end, we consider an $SU(7)$ gauge theory with two anti-symmetric tensors, $A$, and six anti-fundamentals, $\tilde Q$ \CsakiZB. This theory has a one parameter family of non-anomalous $R$-symmetries given by $\CR^t(\tilde Q)=y, \ \CR^t(A)={1\over5}(1-3y)$. Here, $y$ parameterizes mixing with the conserved, non-$R$ symmetry, $J$, under which $J(A)=3$ and $J(\tilde Q)=-5$. The trial $U$-charges are $U^t(\tilde Q)={3y\over2}-1, \ U^t(A)=-{1\over10}(7+9y)$, and so the trial two point function coefficient in the UV is
\eqn\kUUVSUvii{
\tau_U^{UV,t}={21\over50}(149-174y+306y^2)~.
}

The IR is described by thirty $A\tilde Q^2$ composites and eighteen $A^4\tilde Q$ composites. These degrees of freedom transform in the ${\bf 2\times15}$ and ${\bf 3\times 6}$ representations of the $SU(2)\times SU(6)$ global symmetry. Furthermore, these fields have $R$-charges $\CR^t(A\tilde Q^2)={1\over5}(1+7y), \ \CR^t(A^4\tilde Q)={1\over5}(4-7y)$ and trial $U$-charges $U^t(A\tilde Q^2)={7\over10}(-1+3y), \ U^t(A^4\tilde Q)={1\over10}(2-21y)$. Therefore, the trial IR two point function coefficient is
\eqn\kIRSUvii{
\tau_U^{IR, t}={3\over50}(257-1722y+3528y^2)~.
}
Note that for $y>{1\over231}\left(42+5\sqrt{1281}\right)$ or $y<{1\over231}\left(42-5\sqrt{1281}\right)$ (i.e., for large mixing between the $R$ current and the conserved flavor current), the trial two point function coefficient would increase from the UV to the IR. However, $a$-maximization selects a small enough value for $|y|$.

Indeed, by maximizing
\eqn\amaxSUvii{
a^t=96+3\left(42\left(-{1\over5}(4+3y)\right)^3+42\left(y-1\right)^3\right)-\left(42\left(-{1\over5}(4+3y)\right)+42\left(y-1\right)\right)~,
}
we find $y={1\over42}\left(69-5\sqrt{139}\right)$, and so the resulting UV and IR flavor two point function coefficients are
\eqn\kUVSUvii{
\tau_U^{UV}={3\over7}\left(1266-97\sqrt{139}\right), \ \ \ \tau_U^{IR}=834-{141\sqrt{139}\over2}~.
}
Therefore, we have
\eqn\kconjSUvii{
\tau_U^{UV}>\tau_U^{IR}~,
}
as desired.

Before concluding this section, let us note that this example illustrates why we do not use unitarity bounds in determining $\CR_{\rm vis}$. Indeed, in this example such a procedure would not define an $\CR_{\rm vis}$ symmetry. To see this, note that all the IR operators have $R$-charges less than $2/3$ with respect to the $R$ symmetry determined by \amaxSUvii. Therefore, as explained in section 2, the deformation in \auvdef\ would lead to a $y$-independent trial $a^t$ charge (equal to the correct free $a_{IR}$ charge).

\subsec{An aside on irrelevant deformations}
Note that we can in principle turn on various irrelevant deformations of the free UV fixed point in addition to the marginally relevant gauge coupling. For example, we can consider adding an irrelevant $R$-symmetric superpotential
\eqn\irrW{
W=\lambda \sum\left(A\tilde Q^2\right)_{i_bI_bJ_b}^a\left(A^4\tilde Q\right)^{a+1}_{j_bk_bK_b}+\cdot\cdot\cdot~,
}
where the sum is over the $i_b, j_b, k_b$ $SU(2)$ indices and the $I_b, J_b, K_b$ $SU(6)$ indices (here $b=1,\cdot\cdot\cdot, a$). The ellipses in \irrW\ are for any additional terms that preserve the same $R$ symmetry. For $a\ge3$, we find that the $R$-symmetry preserved under the deformation \irrW\ has $y>{1\over231}\left(42+5\sqrt{1281}\right)$. Therefore, in this case, it is not clear why $\tau_U^{UV}>\tau_U^{IR}$.

The main point is that the theory with the deformation in \irrW\ arises as the effective description of some other theory, which we denote as $\CT$. In particular, we can define $\tau_U^{UV}$ for the free fixed point of anti-fundamentals and anti-symmetric tensors only when the RG flow from $\CT$ passes arbitrarily close to it. In this case, there is an arbitrarily-well conserved $U^{UV}_{\mu}$ operator and a corresponding $\tau_U^{UV}$. Since the deformation in \irrW\ is irrelevant, this deformation remains arbitrarily small as we flow into the deep IR, and we find that the $R$ symmetry defined in the previous section is arbitrarily-well preserved along the flow from the free theory of anti-fundamentals and anti-symmetric tensors to the IR fixed point built out of composites. Therefore, we can take $\tau_U^{UV}$ for this theory to be the same as the one in the previous section, and so it is larger than $\tau_U^{IR}$.

Thus, we see that $\tau_U$ is piecewise-defined for the flow from $\CT$ to the free theory of anti-fundamentals and anti-symmetric tensors and then for the flow from this fixed point to the IR theory of free composites. The above argument breaks down when we consider turning on dangerously-irrelevant deformations, since these deformations become relevant in the IR. Indeed, we will need to include these deformations when we determine $\CR_{\rm vis}^{UV}$. However, as we will see in the appendices, our conjecture seems to hold in theories with dangerously irrelevant operators as well.

\newsec{The theory of Intriligator, Seiberg, and Shenker}
In the two previous sections, we tested our conjecture in some well-understood examples. In the appendices we continue this study in a variety of different theories, including interacting SCFTs with accidental symmetries.

 In this section, however, we would like to apply our criterion to a theory whose IR phase is more mysterious. In particular, we will study the case of the $SU(2)$ gauge theory with a single field, $Q$, in the isospin $3/2$ representation. This theory was studied in \IntriligatorRX, where the authors argued that if the theory admitted an IR confining description in terms of a single chiral gauge invariant operator, $u=Q^4$, one could deform the theory with a superpotential of the form $W=\lambda u$ and find a very simple model of dynamical SUSY breaking.

While the authors of \IntriligatorRX\ could not definitively determine the IR phase of the theory, they noted that since there is a non-anomalous $R$ symmetry under which $\CR_{\rm vis}^{UV}(Q)=3/5$ and $\CR_{\rm vis}^{IR}(u)=12/5$ (before deforming the theory), the $U(1)_R$ and $U(1)_R^3$ anomalies matched. Therefore, Intriligator, Seiberg, and Shenker were led to conjecture that the theory was confining in the IR. Subsequently, new evidence has pointed to the opposite conclusion---namely, that the theory in the IR is in fact interacting conformal \refs{\IntriligatorIF, \PoppitzKZ, \VartanovXJ}.

As we will see, our criterion also favors the interpretation of the IR as being interacting conformal. Indeed, note that $U^{UV}_{\rm vis}(Q)=-{1\over10}, \ U^{IR}_{\rm vis}(u)={13\over5}$.
Therefore
\eqn\tauconj{
\tau_U^{UV}={1\over25}, \ \ \ \tau_U^{IR, \rm confining}={169\over25}~.
}
In particular, we find that
\eqn\tauconjviol{
\tau_U^{UV}<\tau_U^{IR, \rm confining}~,
}
which would violate our conjecture by a large margin. Therefore, we conclude that if our conjecture holds, the IR theory is in an interacting conformal phase---in agreement with \refs{\IntriligatorIF, \PoppitzKZ, \VartanovXJ}---and that the IR SCFT has at most a very small amount of mixing between any accidental symmetries and the IR superconformal $R$ current.

Note that since this theory is just barely asymptotically free (the one-loop beta function coefficient is $b=6-5=1$), our conjecture formalizes the intuition that the theory should not be strongly coupled enough to produce confined degrees of freedom. We will test these ideas again in the appendices in the better-understood misleading anomaly matching theory of Brodie, Cho, and Intriligator \BrodieVV.

\newsec{Discussion and conclusions}
By its nature, our paper leaves many open questions. First and foremost, can we prove that $\tau_U^{UV}>\tau_U^{IR}$ for theories with accidental symmetries? A direct proof of this statement by studying the spectral properties of the $\langle U^{IR}(x)U^{IR}(0)\rangle$ two point function is rather difficult since the spectral density of this correlator has infinite support in an SCFT with accidental symmetries (i.e., it is not like the delta function spectral density for $\langle T_{\mu\nu}(x)T_{\rho\lambda}(0)\rangle$ which was exploited in the dispersive proof of the $c$-theorem \CappelliYC). 

This non-compact support implies that we cannot simply use positivity of the spectral density to show that $\tau_U^{UV}-\tau_U^{IR}>0$. This is just as well since otherwise we would find that flavor currents of the full theory also have decreasing two point functions---a statement which we know to be false in general. Of course, our $U$ operator is related by SUSY to the trace of the stress tensor away from the fixed points, but even this fact does not overcome the impediments mentioned above (as one can see by simple dimensional analysis on the form the spectral density must take in the SCFT with accidental symmetries).

In this naive analysis, we have not used the fact that our definition of $\tau_U$ is related to maximizing $a$ in the deformed UV theory. Somehow we must make contact with this fact at the level of the spectral density. Alternatively, it may be useful to note that our conjecture can be rephrased as follows
\eqn\adefconj{
\partial^2_{t^{UV}_U}a^t_{UV}|_{t^i=t_*^{i, UV}}<\partial^2_{t^{IR}_U}a^t_{IR}|_{t^i=t_{*}^{i, IR}}~.
}
One may then hope that we can make contact with the recent $a$-theorem proof in \KomargodskiVJ. Indeed, it may be that the trial $a$ function manifests itself in the dilaton effective action, and so one might be able to make progress. We hope to return to these questions shortly.

Before concluding, we should also note that $\tau_U$ has a natural extension to three dimensions with $Z$-extremization \JafferisUN\ playing the role of $a$-maximization in the corresponding three dimensional quantity (the relevant SUSY multiplets were discussed recently in \DumitrescuIU).  We have not checked if this extended conjecture holds in specific three dimensional examples, but this may be worth doing.

Finally, let us note that we could also consider a different operator, ${U'}_{\rm vis}^{UV}$, and corresponding two-point function coefficient, $\tau_U'$, whose definitions do not depend on $a$-maximization. Indeed, we could simply define ${U'}^{UV}_{\mu, \rm vis}$ to satisfy \seconddem\ for all symmetries of the full RG flow, i.e. $\langle {U'}^{UV}_{\mu, \rm vis}(x)\hat J^{UV*}_{\nu, a}(0)\rangle=0$ for all $a$. In fact, it turns out that ${\tau'}_U^{UV}>{\tau'}_U^{IR}$ in all of the examples we have explored above (note that $\tau_U\ne{\tau'}_{U}$ for the theory in section 4). Furthermore, ${\tau'}_U^{UV}$ has the advantage of being simpler to define than $\tau_U^{UV}$ (although, in many examples, $U_{\rm vis}=U_{\rm vis}'$). Its main disadvantage, however, is that unlike $\tau_U$, it is not immediately clear that ${\tau'}_U^{UV}>{\tau'}_U^{IR}$ in the class of theories without accidental symmetries. Clearly, it is worth investigating more thoroughly if ${\tau'}_U^{UV}>{\tau'}_U^{IR}$ as well. 

\smallskip\smallskip\smallskip\centerline{\bf Acknowledgements}~
We are grateful to S. Abel, I. Antoniadis, Z. Komargodski, and N. Seiberg for interesting comments and discussions. We would also like to acknowledge Z. Komargodski for discussions that led to the result in section 5. This work was supported in part by the European Commission under the ERC Advanced Grant 226371 and the contract PITN-GA-2009-237920.

\vfill\eject
\appendix{A}{$SO(N_c)$ SQCD}
In this appendix, we will consider an $SO(N_c)$ gauge theory with $N_f$ quarks, $Q$, transforming in the vector representation \refs{\SeibergPQ, \IntriligatorID, \IntriligatorAU}. This theory is asymptotically free for $N_f<3(N_c-2)$. In this case, there is a unique visible $\CR_{\rm vis}$ symmetry in the UV $\CR_{\rm vis}^{UV}(Q)={N_f-(N_c-2)\over N_f}$, and so $U_{\rm vis}^{UV}(Q)={6-3N_c+N_f\over2N_f}$. For $N_f\le N_c-5$, this theory has no ground state.

\subsec{Confinement: $N_f=N_c-4$}
In this phase, we find
\eqn\kuvconfSO{
\tau_U^{UV}={N_c(N_c-1)^2\over N_c-4}~.
}
In the IR we have a theory of ${1\over2}(N_c-4)(N_c-3)$ mesons, $M$. These operators have $\CR_{\rm vis}$-charge $\CR^{IR}_{\rm vis}(M)={4\over4-N_c}$, and $U$-charge $U_{\rm vis}^{IR}(M)={2+N_c\over 4-N_c}$. Therefore, we see that
\eqn\kUIconfSO{
\tau_U^{IR}={(N_c-3)(2+N_c)^2\over 2(N_c-4)}~,
}
and it follows that our conjecture holds in this case
\eqn\dkconfSO{
\tau_U^{UV}>\tau_U^{IR}~.
}

\subsec{Confinement: $N_f=N_c-3$}
Here we find
\eqn\kuvconfiiSO{
\tau_U^{UV}={N_c(2N_c-3)^2\over 4(N_c-3)}~.
}
The IR spectrum consists of ${1\over2}(N_c-3)(N_c-2)$ mesons, $M$, and $N_c-3$ singlets, $q$. These operators have charges $\CR_{\rm vis}^{IR}(M)={2\over 3-N_c}, \ \CR^{IR}_{\rm vis}(q)=1+{1\over N_c-3}$,
and $U_{\rm vis}^{IR}(M)={N_c\over 3-N_c}, \ U_{\rm vis}^{IR}(q)={N_c\over2(N_c-3)}$.
Therefore, we have
\eqn\kIRconfiiSO{
\tau_U^{IR}={(2N_c-3)N_c^2\over 4(N_c-3)}~.
}
It is straightforward to check that
\eqn\dkconfiiSO{
\tau_U^{UV}>\tau_U^{IR}~,
}
as desired.

\subsec{Abelian Coulomb phase: $N_f=N_c-2$}
In this phase we have
\eqn\kUVabco{
\tau_U^{UV}=N_c(N_c-2)~.
}
The low energy theory is in an abelian Coulomb phase with a description in terms of ${1\over2}(N_c-2)(N_c-1)$ mesons, $M$, and $N_f$ monopole pairs, $q_{\pm}$. These operators have charges $\CR^{IR}_{\rm vis}(M)=0, \ \CR^{IR}_{\rm vis}(q^{\pm})=1$, and $U_{\rm vis}^{IR}(M)=-1, \ U_{\rm vis}^{IR}(q^{\pm})={1\over2}$.
From this discussion, we see that
\eqn\kIRmono{
\tau_U^{IR}={1\over2}N_c(N_c-2)~,
}
and so our conjecture holds in this phase
\eqn\kconfmono{
\tau_U^{UV}>\tau_U^{IR}~.
}

Recall that there is also a submanifold of the moduli space that is in a free dyonic phase (with dyons $E^{\pm}$). The dyons have charges $\CR_{\rm vis}(E^{\pm})=1$ and $U_{\rm vis}^{IR}(E^{\pm})={1\over2}$.
As a result, 
\eqn\taudyon{
\tau_U^{IR}={1\over2}(3-3N_c+N_c^2)~,
}
and we again see that our conjecture is confirmed
\eqn\dyondk{
\tau_U^{UV}>\tau_U^{IR}~.
}

\subsec{The free magnetic phase: $N_c-2< N_f\le 3(N_c-2)/2$}
In this phase we find
\eqn\kUVfmSO{
\tau_U^{UV}={N_c(6-3N_c+N_f)^2\over 4N_f}~.
}
The IR is described by a free magnetic theory with ${1\over2}N_f(N_f+1)$ singlet mesons, $M$, and $N_f$ quarks, $q$, transforming as vectors of the dual $SO(N_f-N_c+4)$ gauge group. The charges of these fields are $\CR_{\rm vis}^{IR}(M)={2(2-N_c+N_f)\over N_f}, \ \CR_{\rm vis}^{IR}(q)={N_c-2\over N_f}$, and $U_{\rm vis}^{IR}(M)={6-3N_c+2N_f\over N_f}, \ U_{\rm vis}^{IR}(q)=-1+{3(N_c-2)\over 2N_f}$.
Therefore, we find
\eqn\kIRfmSO{
\tau_U^{IR}={(6-N_c+3N_f)(6-3N_c+2N_f)^2\over 4N_f}~.
}
It is straightforward to check that in the free magnetic range our conjecture is satisfied
\eqn\deltakfmSO{
\tau_U^{UV}>\tau_U^{IR}~.
}
This inequality breaks down in a range around $N_f\sim 1.79(N_c-2)$, which is safely within the conformal window (although it is not too far from the boundary).

\subsec{The conformal window: $3(N_c-2)/2< N_f< 3(N_c-2)$}
Just as in the case of the conformal window of $SU(N_c)$ SQCD, we assume there are no accidental symmetries and so
\eqn\kconfSO{
\tau_U^{UV}>0=\tau_U^{IR}~.
}

\subsec{Relevant deformations in the conformal window}
In this section, we repeat the analysis of relevant deformations of $SU(N_c)$ SQCD in the theory at hand. Again, an interesting case to check is to start from an interacting conformal fixed point of the $SO(N_c)$ theory and turn on the following deformation
\eqn\massSOint{
W=\lambda Q_aQ^a~,
}
where $a=1,...,k$. In the case that $k<N_f-{3\over2}(N_c-2)$, the theory flows to another fixed point in the conformal window and $\tau_U^{UV}>0=\tau_U^{IR}$ (to define $\tau_U^{UV}$ for any $k>0$, recall the discussion around \seconddem). In the case that $k=N_f-{3\over2}(N_c-2)$, we again find $\tau_U^{UV}>0=\tau_U^{IR}$ since $\CR_{\rm vis}$ flows to the free superconformal $R$ current.

Next, consider the regime $N_f-{3\over2}(N_c-2)\le k< N_f-N_c+2$. We find that the coefficient of the two point function of $U_{\mu}$ in the interacting conformal theory is
\eqn\tauintUVSO{
\tau_U^{UV}={27kN_c(N_c-2)^3\over 4(N_f-k)N_f^2}~.
}
The corresponding quantity at the free magnetic fixed point is
\eqn\taufmSOint{
\tau_U^{IR}={(6-N_c+3(N_f-k))(6-3N_c+2(N_f-k))^2\over4(N_f-k)}~,
}
and so it is straightforward to show that
\eqn\tauconjintSOfm{
\tau_U^{UV}> \tau_U^{IR}~.
}

In the cases that $k=N_f-N_c+2$, $k=N_f-N_c+3$, and $k=N_f-N_c+4$, we use \tauintUVSO\ for $\tau_U^{UV}$ and the results of the relevant previous subsections for $\tau_U^{IR}$. It is simple to conclude that
\eqn\tauconjintSOfm{
\tau_U^{UV}> \tau_U^{IR}~,
}
thus proving our conjecture for this class of RG flows.

\subsec{Higgsing}
In this section we consider RG flows in $SO(N_c)$ SQCD that involve Higgsing. We can again imagine starting in the asymptotically free limit and turning on a set of vevs $\langle Q^a_a\rangle =v_a$ with all the $v_a$ ($a=1,...,k$) distinct. This Higgsing breaks $SO(N_c)\to SO(N_c-k)$ and breaks the flavor symmetry $SU(N_f)\to SU(N_f-k)$. We then find a theory of $SO(N_c-k)$ SQCD with $N_f-k$ fundamentals, $\CQ$, and a set of singlet modes. These latter fields are characterized by a set of $k$ gauge singlets, $\Phi_a$, transforming in the ${\bf N_f-k}$ representation of the unbroken flavor symmetry and a set of $k(k+1)/2$ singlets, $S$. We can define an $\CR_{\rm vis}$ symmetry by demanding that it leave the vacuum invariant. We then find that $\CR_{\rm vis}^{IR}(\CQ)=\CR_{\rm vis}^{IR}(\Phi)=1-{N_c-2-k\over N_f-k}, \ \CR_{\rm vis}^{IR}(S)=0$ (for $k=N_c-2, N_c-1, N_c$ we must recall the discussion around \seconddem). Again, by a similar discussion to the one in the $SU(N_c)$ case, Higgsed RG flows starting from the asymptotically free limit satisfy $\tau_U^{UV}>\tau_U^{IR}$.

Let us now consider Higgsing a fixed point in the conformal window. If $k<(3(N_c-2)-N_f)/2$, then the theory flows to a new interacting fixed point in the IR with a set of decoupled, free singlet modes of the type described in the previous paragraph. In the UV, we then find
\eqn\tauUVhiggsSO{
\tau_U^{UV}={27k(N_c-2)N_c(2-N_c+N_f)^2\over 4(N_f-k)N_f^2}~.
}
On the other hand, in the IR, we see that
\eqn\tauIRhiggsSO{
\tau_U^{IR}={k(2k^2+N_f(2+N_f(1-3N_c/N_f)^2)-2k(1+N_f(-3+6N_c/N_f)))\over 4(N_f-k)}~.
}
It is then easy to see that
\eqn\tauconjhiggsSO{
\tau_U^{UV}>\tau_U^{IR}~,
}
as desired.

If, on the other hand $(3(N_c-2)-N_f)/2\le k\le N_c$, the IR fixed point is free. The UV two point function coefficient is as in \tauUVhiggsSO, but in the IR we now have
\eqn\tauirhiggsSOfree{\eqalign{
\tau_U^{IR}&=-{1\over4(N_f-k)}\Big(2k^3-N_f^2(1-3N_c/N_f)^2(2+N_c)-2k^2(3+N_f+2N_c)\cr&+2kN_f(-5-2(-6+N_f)N_c/N_f+6N_c^2/N_f)\Big)~.
}}
Therefore, 
\eqn\tauconjhiggsIRfreeSO{
\tau_U^{UV}>\tau_U^{IR}~.
}

\appendix{B}{$Sp(N_c)$ SQCD}
In this section we consider $Sp(N_c)$ SQCD with $2N_f$ quarks, $Q$, in the fundamental ($2N_c$ dimensional) representation \IntriligatorNE. The theory is asymptotically free for $N_c<3(N_c+1)$ and, much like $SU$ and $SO$ SQCD has a unique $\CR_{\rm vis}$ symmetry with $\CR_{\rm vis}^{UV}(Q)=1-{N_c+1\over N_f}$, and a corresponding $U_{\rm vis}^{UV}$-charge $U_{\rm vis}^{UV}(Q)={-3-3N_c+N_f\over 2N_f}$. This theory does not have a vacuum for $N_f\le N_c$.

\subsec{Deformed moduli space: $N_f=N_c+1$}
In this phase we find that in the UV
\eqn\kUVdefmSP{
\tau_U^{UV}=4N_c(1+N_c)~.
}
The IR is described by the $(N_c+1)(2N_c+1)$ mesons, $M$, subject to the constraint
\eqn\defmodSP{
{\rm Pf} M\sim\Lambda^{2(N_c+1)}~.
}
The mesons have $\CR_{\rm vis}^{IR}$ charge $\CR_{\rm vis}^{IR}(M)=0$ and corresponding $U_{\rm vis}^{IR}$-charge $U_{\rm vis}^{IR}(M)=-1$. We then find
\eqn\kUIRdmSP{
\tau_U^{IR}<(1+N_c)(1+2N_c)~,
}
since the deformed moduli space constraint removes some of the degrees of freedom due to the quantum constraint \defmodSP. Therefore, our conjecture is compatible with this result
\eqn\deltakdmSP{
\tau_U^{UV}>\tau_U^{IR}~.
}

\subsec{$s$-confinement: $N_f=N_c+2$}
In this phase, we have
\eqn\kUVsconf{
\tau_U^{UV}={N_c(1+2N_c)^2\over2+N_c}~.
}
The IR is described by $(N_c+2)(2N_c+3)$ mesons, $M$. These fields have $\CR_{\rm vis}$-charge $\CR_{\rm vis}^{IR}={2\over N_c+2}$, and $U$-charges $U_{\rm vis}^{IR}(M)={1-N_c\over 2+N_c}$. We then find
\eqn\kIRscSP{
\tau_U^{IR}={(3+2N_c)(N_c-1)^2\over2+N_c}~,
}
from which it is straightforward to see that our conjecture holds
\eqn\kconjscSP{
\tau_U^{UV}>\tau_U^{IR}~.
}

\subsec{The free magnetic phase: $N_c+2< N_f\le3(N_c+1)/2$}
In this phase, the UV two point function coefficient is
\eqn\kUVfmSP{
\tau_U^{UV}={N_c(-3-3N_c+N_f)^2\over N_f}~.
}
The IR is described by a dual $Sp(N_f-N_c-2)$ gauge theory with $N_f(2N_f-1)$ mesons, $M$, and $2N_f$ squarks, $q$, in the fundamental representation. These operators have $\CR_{\rm vis}$-charges $\CR_{\rm vis}^{IR}(M)=2\left(1-{1+N_c\over N_f}\right), \  \CR_{\rm vis}^{IR}(q)={1+N_c\over N_f}$. The resulting $U$-charges are $U(M)={-3-3N_c+2N_f\over N_f}, \ U(q)=-1+{3(1+N_c)\over 2N_f}$. We then see
\eqn\kUIRfmSP{
\tau_U^{IR}={(-3-N_c+3N_f)(3+3N_c-2N_f)^2\over N_f}~.
}
It is then simple to verify that for the free magnetic range
\eqn\dkfmSP{
\tau_U^{UV}>\tau_U^{IR}~.
}
This inequality breaks down at $N_f\sim1.79(N_c+1)$, which is safely inside the conformal window (but still close to the boundary).

\subsec{Conformal window: $3(N_c+1)/2<N_f<3(N_c+1)$}
Just as in the $SO$ and $SU$ cases discussed above, our conjecture follows in this case from the (assumed) lack of accidental symmetries in the $Sp$ conformal window. Therefore
\eqn\kconfwinSP{
\tau_U^{UV}>\tau_U^{IR}~.
}

\subsec{Relevant deformations in the conformal window}
As in the $SU$ and $SO$ SQCD theories, we would like to check that deforming an interacting $Sp(N_c)$ SQCD fixed point by a relevant deformation is compatible with our conjecture. To that end, consider deforming the theory by
\eqn\WsPint{
W=\lambda Q_{2a-1}Q_{2a}~,
}
where $a=1,...,k$. For $k<N_f-{3\over2}(N_c+1)$, the theory flows to another interacting conformal fixed point and so $\tau_U^{UV}>0=\tau_U^{IR}$ (to define $\tau_U^{UV}$ for any $k>0$, recall the discussion around \seconddem). In the case that $k=N_f-{3\over2}(N_c+1)$, the theory flows to a free theory with $\tau_U^{UV}>0=\tau_U^{IR}$ as well.

Next, we consider taking $N_f-{3\over2}(N_c+1)\le k< N_f-N_c-2$. At the interacting fixed point, we recall the discussion around \seconddem\ and find
\eqn\tauUVSpint{
\tau_U^{UV}={27k N_c(N_c+1)^3\over (N_f-k)N_f^2}~,
}
while at the free magnetic fixed point we have
\eqn\tauIRfmSP{
\tau_U^{IR}={(-3-N_c+3(N_f-k))(3+3N_c-2(N_f-k))^2\over N_f-k}~.
}
It is straightforward to verify that
\eqn\tauconjSpintfm{
\tau_U^{UV}> \tau_U^{IR}~.
}

Finally, consider taking $k=N_f-N_c-2$ and $k=N_f-N_c-1$. Using \kIRscSP\ and \kUIRdmSP\ respectively, it is simple to check that
\eqn\tauconjSpintfm{
\tau_U^{UV}> \tau_U^{IR}~,
}
for this set of parameters as well.

\subsec{Higgsing}
In this section we consider RG flows with Higgsing in the $Sp(N_c)$ series of theories. The discussion proceeds much as it did for $SU$ and $SO$ SQCD. We can imagine starting in the asymptotically free regime and turning on squark vevs, $\langle Q^{2a-1}_{2a-1}\rangle=\langle Q^{2a}_{2a}\rangle=v_a$ (where $a=1,...,k$), with all the $v_a$ distinct. The Higgsing breaks $Sp(N_c)\to Sp(N_c-k)$ and $SU(2N_f)\to Sp(1)^k\times SU(2(N_f-k))$. We then find a sector that consists of an $Sp(N_c-k)$ SQCD with $2(N_f-k)$ fundamentals, $\CQ$, and a sector of $2k$ gauge singlets, $\Phi$, transforming in the fundamental representation of $SU(2(N_f-k))$ as well as a set of $k(2k-1)$ singlets, $S$. We can define an $\CR_{\rm vis}$ symmetry by demanding that it leaves the vacuum invariant. We then find $\CR_{\rm vis}^{UV}(\CQ)=\CR_{\rm vis}^{UV}(\Phi)=1-{N_c-k+1\over N_f-k}, \ \CR_{\rm vis}^{UV}(S)=0$. By a similar discussion to the one in the $SU$ and $SO$ case, Higgsed RG flows starting from the asymptotically free limit satisfy $\tau_U^{UV}>\tau_U^{IR}$.

Let us now consider Higgsing an interacting conformal fixed point. As long as $k<{\rm min}\left((3(N_c+1)-N_f)/2, N_c\right)$, the theory flows to another fixed point in the conformal window. In the UV, we find
\eqn\tauUVSphiggs{
\tau_U^{UV}={27k N_c(N_c+1)(1+N_c-N_f)^2\over (N_f-k)N_f^2}~.
}
In the IR, we find
\eqn\tauIRSphiggs{
\tau_U^{IR}={k\over N_f-k}\left(9+2k^2+9N_c^2-6N_c(N_f-3)-7N_f+N_f^2+k(-11-12N_c+6N_f)\right)~.
}
Therefore,
\eqn\tauconjHiggsint{
\tau_U^{UV}>\tau_U^{IR}~,
}
as desired.

Next, let us consider the case that $(3(N_c+1)-N_f)/2\le k\le N_c$. We again use \tauUVSphiggs\ for the UV two point function coefficient, but now in the IR, we have
\eqn\tauconIRSphiggsf{\eqalign{
\tau_U^{IR}&={1\over N_f-k}\Big(-2k^3+N_c(-3-3N_c+N_f)^2-k(12N_c^2-4N_c(-3+N_f)+N_f)\cr&+k^2(1+4N_c+2N_f)\Big)~,
}}
and so
\eqn\tauconjHiggssp{
\tau_U^{UV}>\tau_U^{IR}~.
}

\appendix{C}{More free magnetic theories}
In this section we will consider two more complicated theories in their free magnetic range. We start in the next subsection by analyzing the Kutasov theory and conclude with an analysis of the Brodie theory in the final subsection. These theories also illustrate the important role played by dangerously irrelevant operators.

\subsec{The Kutasov theory}
In this section we will consider adding an adjoint field, $X$, to $SU(N_c)$ SQCD. We will study the theory with the superpotential
\eqn\WXKut{
W=s_0\Tr X^{k+1}~,
}
where we take $k\ge2$ \refs{\KutasovVE\KutasovXU-\KutasovSS}. The deformation \WXKut\ is dangerously irrelevant at the free UV fixed point and selects a unique $\CR_{\rm vis}$ symmetry under which the various UV fields transform as $\CR_{\rm vis}^{UV}(Q)=\CR_{\rm vis}^{UV}(\tilde Q)=1-{2N_c\over(1+k)N_f}, \ \CR_{\rm vis}^{UV}(X)={2\over k+1}$. The corresponding $U$-charges are $U_{\rm vis}^{UV}(Q)=U_{\rm vis}^{UV}(\tilde Q)={1\over2}-{3N_c\over(1+k)N_f}, \ U_{\rm vis}^{UV}(X)=-1+{3\over k+1}$. Therefore, the $U$ two point function coefficient in the UV is
\eqn\kUVkut{
\tau_U^{UV}={36N_c^3-2(k-2)^2N_f+2(k^2-10k-2)N_fN_c^2+(1+k)^2N_cN_f\over2N_f(1+k)^2}~.
}

For ${N_c\over k}<N_f\le{2N_c\over2k-1}$, the IR theory is in a free magnetic phase described by an $SU(kN_f-N_c)$ gauge theory with mesons $M_j=QX^{j-1}\tilde Q$ ($j=1,..., k$), an adjoint, $Y$, and $N_f$ flavors of dual quarks, $q$ and $\tilde q$. These fields have $\CR_{\rm vis}$-charge $\CR_{\rm vis}^{IR}(M_j)=2\left(1-{2N_c\over(1+k)N_f}\right)+{2(j-1)\over k+1}, \ \CR_{\rm vis}^{IR}(q)=\CR_{\rm vis}^{IR}(\tilde q)=1-{2(kN_f-N_c)\over(1+k)N_f},\ \CR_{\rm vis}^{IR}(Y)={2\over k+1}$, and corresponding $U$-charges $U_{\rm vis}^{IR}(M)=2-{6N_c\over(k+1)N_f}+{3(j-1)\over k+1}, \  U_{\rm vis}^{IR}(q)=U_{\rm vis}^{IR}(\tilde q)={6N_c+(1-5k)N_f\over2(1+k)N_f}, \ U_{\rm vis}^{IR}(Y)=-1+{3\over k+1}$,
from which we conclude
\eqn\kIRkut{\eqalign{
\tau_U^{IR}&={1\over2N_f(1+k)^2}\Big(-36N_c^3+2(-2+80k+k^2)N_fN_c^2-(1+6k+153k^2+4k^3)N_cN_f^2\cr&+(-8+8k+43k^3N_f^2+2k^4N_f^2+(-2+5N_f^2)k^2)N_f\Big)~.
}}
It is easy to check that in the free magnetic range
\eqn\kconjKut{
\tau_U^{UV}>\tau_U^{IR}~.
}

\subsec{The Brodie theory}
In this subsection we study $SU(N_c)$ SQCD with two adjoints---$X$ and $Y$---and the following dangerously-irrelevant superpotential introduced in the free UV theory \BrodieVX
\eqn\WXY{
W=s_0\Tr X^{k+1}+\Tr XY^2~,
}
with $k\ge2$. The above superpotential fixes the $\CR_{\rm vis}$ charges of the UV fields as follows $\CR_{\rm vis}^{UV}(Q)=\CR_{\rm vis}^{UV}(\tilde Q)=1-{N_c\over(1+k)N_f}, \ \CR_{\rm vis}^{UV}(X)={2\over k+1}, \ \CR_{\rm vis}^{UV}(Y)={k\over k+1}$. The corresponding $U$-charges are $U_{\rm vis}^{UV}(Q)=U_{\rm vis}^{UV}(\tilde Q)={1\over2}-{3N_c\over2(1+k)N_f}, \ U_{\rm vis}^{UV}(X)=-1+{3\over k+1}, \ U_{\rm vis}^{UV}(Y)={k-2\over2( k+1)}$. Therefore, in the UV, we have
\eqn\kUVBr{
\tau_U^{UV}={18N_c^3-5(k-2)^2N_f+(8-32k+5k^2)N_fN_c^2+2(k+1)^2N_cN_f^2\over4N_f(1+k)^2}~.
}
For ${N_c\over3k}<N_f\le{N_c\over3k-1}$ the theory is in a free magnetic phase with gauge group $SU(3kN_f-N_c)$ and $3k$ mesons, $M_{lj}=QX^{j-1}Y^{l-1}\tilde Q$ ($l=1,2,3$ and $j=1,...,k$), two adjoints $\tilde X, \tilde Y$, and $N_f$ flavors $q, \tilde q$. The $\CR_{\rm vis}$-charges are $\CR_{\rm vis}^{IR}(M_{lj})={-2N_c+(2j+k+k l)N_f\over(1+k)N_f}, \ \CR_{\rm vis}^{IR}(q)=\CR_{\rm vis}^{IR}(\tilde q)=1-{3k N_f-N_c\over N_f(k+1)}, \ \CR_{\rm vis}^{IR}(\tilde X)={2\over k+1}, \ \CR_{\rm vis}^{IR}(\tilde Y)={k\over k+1}$, while the $U$-charges are $U_{\rm vis}^{IR}(M_{lj})=-1+{3(-2N_c+(2j+k+kl)N_f)\over2(1+k)N_f}, \ U_{\rm vis}^{IR}(q)=U_{\rm vis}^{IR}(\tilde q)={3N_c+N_f-8kN_f\over 2(1+k)N_f},  \ U_{\rm vis}^{IR}(\tilde X)=-1+{3\over k+1}, \ \ \ U(\tilde Y)={k-2\over2( k+1)}$.
Therefore, we have
\eqn\kUIRBr{\eqalign{
\tau_U^{IR}&={1\over4(1+k)^2N_f}(-18N_c^3+(8+238k+5k^2)N_fN_c^2-2(1+44k+328k^2+15k^3)N_cN_f^2\cr&+N_f(-20+20k+531k^3N_f^2+45k^4N_f^2+k^2(-5+144N_f^2)))~.
}}
It is again easy to check that
\eqn\conjkBr{
\tau_U^{UV}>\tau_U^{IR}~.
}

\appendix{D}{$\CN=2$ $SU(N_c)$ Super Yang-Mills (SYM)}
Up to now, all of our examples have been in $\CN=1$ SUSY. In this section, we will consider $SU(N_c)$ $\CN=2$ SYM \refs{\SeibergRS\DouglasNW\ArgyresXH\KlemmQS-\KlemmQJ}. The UV consists of an adjoint field, $\Phi$, with $\CR_{\rm vis}^{UV}(\Phi)=0$ and $U_{\rm vis}^{UV}(\Phi)=-1$. Therefore,
\eqn\kUVSYM{
\tau_U^{UV}=N_c^2-1~.
}

In the IR, we check the $N_c$ vacua on the Coulomb branch where $N_c-1$ monopoles / dyons become massless. The IR $\CR_{\rm vis}$ symmetry is fixed by the $SU(2)_R$ symmetry of the theory. Indeed, it simply corresponds to the $I_3\subset SU(2)_R$ generator. Since the monopoles are hypermultiplets, their charges are then $\CR_{\rm vis}^{IR}(E)=\CR_{\rm vis}^{IR}(\tilde E)=1$ (we also have that $\CR_{\rm vis}^{IR}(\Phi)=0$). Therefore, $U_{\rm vis}^{IR}(E)=U_{\rm vis}^{IR}(\tilde E)={1\over2}, \ U_{\rm vis}^{IR}(\Phi)=-1$, and so
\eqn\kUIR{
\tau_U^{IR}={3\over2}(N_c-1).
}
As a result,
\eqn\kconjNii{
\tau_U^{UV}>\tau_U^{IR}~.
}

Before concluding, note that the $N_c=2$ results match the results in \kIRmono\ and \taudyon\ for $SO(3)$ with $N_f=1$ (as they should).

\appendix{E}{More $s$-confining examples}
In this section, we would like to study our conjecture in some more complicated $s$-confining theories \CsakiZB. Unlike some of the theories discussed above, the theories in this section have a continuous family of candidate $\CR_{\rm vis}$-symmetries, and so we must use the procedure described in the introduction to fix this ambiguity.

\subsec{$SU(5)$ with $3\times({ \bf10 \oplus\bar 5})$}
Let us consider a theory with an $SU(5)$ gauge group, three anti-symmetric tensors, $A$, and three anti-fundamentals, $\tilde Q$. There is a family of non-anomalous $R$-symmetries given by $R^t(\tilde Q)=y, \ R^t(A)={1\over9}(2-3y)$,
where $y$ is a parameter that measures the mixing of the $R$ symmetry with the conserved non-$R$ symmetry, $J$, under which $J(\tilde Q)=-3J(A)$. The corresponding trial $U$-symmetry is $U^t(\tilde Q)={3y\over2}-1, \ \ \ U^t(A)=-{1\over6}(4+3y)$, which leads to the following trial two point function coefficient 
\eqn\kUUVSUv{
\tau_U^{UV, t}={5\over12}(68-60y+99y^2)~.
}

The IR degrees of freedom are the nine $A\tilde Q^2$, the twenty-four $A^3\tilde Q$, and the six $A^5$. These operators have the following $R$-charges $R^t(A\tilde Q^2)={2\over9}+{5y\over3}, \ R^t(A^3\tilde Q)={2\over3}, \ R^t(A^5)={5\over9}(2-3y)$, and the following $U$-charges, $U^t(A\tilde Q^2)=-{2\over3}+{5y\over2}, \ U^t(A^3\tilde Q)=0, \ U^t(A^5)={2\over3}-{5y\over2}$. The IR trial flavor two point function coefficient is then
\eqn\kUIRSUv{
\tau_U^{IR, t}={5\over12}(4-15y)^2~.
}
Note that for very large mixing of the conserved flavor current with $R^t$, i.e., for $|y|\gg1$, the trial IR flavor two point function coefficient becomes larger than the trial UV one. However, as we will see momentarily, after performing $a$-maximization in the deformed UV theory, we will arrive at a value of $y$ that is sufficiently small such that our inequality is satisfied.

To see this, perform $a$ maximization over
\eqn\amaxSUv{
a=48+3\left(15(y-1)^3+30\left(-{1\over9}(7+3y)\right)^3\right)-\left(15(y-1)+30\left(-{1\over9}(7+3y)\right)\right)
}
We find that $y={4\over15}$. Therefore, the flavor two point function coefficients in the UV and IR are
\eqn\kSUv{
\tau_U^{UV}={123\over5}, \ \ \ \tau_U^{IR}=0~.
}
As a result,
\eqn\kconjSUv{
\tau_U^{UV}>\tau_U^{IR}~.
}

\subsec{$SU(2N+1)$ with {$\bf N(2N+1)\oplus\bar{N(2N+1)}\oplus3\times(2N+1\oplus\bar{2N+1})$}}
Let us consider the case of an $SU(2N+1)$ gauge theory with an anti-symmetric tensor, $A$, a conjugate anti-symmetric tensor, $\tilde A$, and three flavors, $Q$ and $\tilde Q$. The family of $R$-symmetries is
\eqn\RsymmtSUiiN{
R^t(Q)=R^t(\tilde Q)=y, \ \ \ R^t(A)=R^t(\tilde A)={1-3y\over 2N-1}~,
}
where $y$ parameterizes mixing with the conserved symmetry under which $A, \tilde A$ have charge $-3$ and $Q, \tilde Q$ have charge $2N-1$. The corresponding $U$-charges are
\eqn\UtSUiiN{
U^t(Q)=U^t(\tilde Q)={3y\over2}-1, \ \ \ U^t(A)=U^t(\tilde A)={5-4N_c-9y\over4N_c-2}~,
}
and the trial UV two point function coefficient is
\eqn\kUVSUiiN{
\tau_U^{UV, t}=2(2N+1)\left({3\over4}(2-3y)^2+{N\over (2-4N)^2}(-5+4N+9y)^2\right)~.
}

The IR is described by $9N$ mesons, $Q(A\tilde A)^k\tilde Q$, $3N$ mesons $\tilde A(A\tilde A)^kQ^2$, $3N$ mesons $A(A\tilde A)^k\tilde Q^2$ (with $k=0,..., N-1$), $N-1$ mesons $(A\tilde A)^m$ ($m=1,..., N-1$), three baryon flavors $A^NQ$, $\tilde A^N\tilde Q$, and one baryon flavor $A^{N-1}Q^3$, $\tilde A^{N-1}\tilde Q^3$. These fields have $R$-charges
\eqn\RsymmtSUiiNIR{\eqalign{
&R^t(Q(A\tilde A)^k\tilde Q)=2\left(y+{k(1-3y)\over 2N-1}\right), \ \ \ R^t((A\tilde A)^m)=m\left(2-6y\over2N-1\right)~,\cr& R^t(\tilde A(A\tilde A)^kQ^2)=R^t(A(A\tilde A)^k\tilde Q^2)=2y+\left({(1+2k)(1-3y)\over2N-1}\right)~,\cr& R^t(A^NQ)=R^t(\tilde A^N\tilde Q)=y+{N(1-3y)\over 2N-1}~, \cr&R^t(A^{N-1}Q^3)=R^t(\tilde A^{N-1}\tilde Q^3)={-1+N(1+3y)\over2N-1}~.
}}
The $U$-charges are
\eqn\USUiiNIR{\eqalign{
&U^t(Q(A\tilde A)^k\tilde Q)=-1+3\left(y+{k(1-3y)\over 2N-1}\right), \ \ \ U^t((A\tilde A)^m)=-1+m\left({3-9y\over 2N-1}\right)~,\cr& U^t(\tilde A(A\tilde A)^kQ^2)=U^t(A(A\tilde A)^k\tilde Q^2)={(5+6k-4N)(-1+3y)\over 2-4N}~,\cr& U^t(A^NQ)=U^t(\tilde A^N\tilde Q)={-2+N+(1+N)3y\over-2+4N}~, \cr&U^t(A^{N-1}Q^3)=U^t(\tilde A^{N-1}\tilde Q^3)={1+N(1-9y)\over2-4N}~.
}}
The corresponding trial IR flavor two point function coefficient is
\eqn\kIRtSUiiN{
\tau_U^{UV,t}={1\over2(1-2N)^2}\left(11-36y+27y^2+6N^2(-1+3y)+9N(-1+3y^2)+4N^3(8-45y+81y^2)\right)
}
It is easy to see that in the large $N$ limit $\tau_U^{UV}>\tau_U^{IR}$ since the UV coefficient goes as $N^2$ while the IR coefficient goes as $N$. Let us now check the remaining range of $N$ (with $y$ fixed by maximizing $a$).

We again maximize $a$
\eqn\amaxSUiiN{\eqalign{
a&=2((2N+1)^2-1)+3\left(6(2N+1)(y-1)^3+2N(2N+1)\left({1-3y\over 2N-1}-1\right)^3\right)\cr&-\left(6(2N+1)(y-1)+2N(2N+1)\left({1-3y\over 2N-1}-1\right)\right)~,
}}
and find
\eqn\ysoln{
y={3+18N^2-24N^3+(2N-1)\sqrt{1+N(-7+N(-1+4N)(9+20N))}\over 3+3N(3+4(3-2N)N)}~.
}
It is then straightforward (though tedious) to check that
\eqn\kconjSUiiN{
\tau_U^{UV}>\tau_U^{IR}~,
}
as desired.

\appendix{F}{Interacting SCFTs with accidental symmetries}
In this section we would like to discuss the case of interacting IR SCFTs with accidental symmetries. We will consider adjoint SQCD in the next subsection and the $\hat D$ SCFT in the final subsection.

The precise IR behavior of these theories is not fully understood, but a consistent picture has begun to emerge \refs{\KutasovIY, \IntriligatorMI}. The basic understanding is the following: as one lowers the number of UV flavors relative to the number of colors in these two theories, some chiral gauge invariant operators hit unitarity bounds (with respect to $\CR_{\rm vis}$) and then decouple in the IR and become free fields. The IR is therefore believed to consist of an interacting SCFT module with no accidental symmetries and a separate, free theory of various gauge invariant operators that transform nontrivially under corresponding accidental $U(1)$ flavor symmetries.

The superconformal $R$ symmetry of the interacting SCFT, $\tilde R|_{\rm int}$, is the restriction of the IR descendant of the $R$ current, $\hat R$, determined by maximizing the deformed $a$ of \auvdef. On the other hand, our $\CR_{\rm vis}$ is determined by maximizing the undeformed $a$ and therefore does not generally agree with IR superconformal $R$ current in either the interacting sector or the free sector. This fact implies that there are nontrivial contributions to $\tau_U^{IR}$ from both the free and interacting sectors.

We can compute $\tau_U^{IR}$ by using \kUIRintfree, which we reproduce below for ease of reference
\eqn\kUIRintfreei{
\tau_U^{IR}=-{27\over4}\Tr\ \hat \CR^{UV}(\CR_{\rm vis}^{UV}-\hat \CR^{UV})^2+{27\over4}\Tr|_{\rm free}\ \hat \CR^{IR}(\CR^{IR}_{\rm vis}-\hat \CR^{IR})^2+\Tr|_{\rm free}\left(U^{IR}_{\rm vis}\right)^2~,
}
where the first term is computed via anomaly matching, and the second and third terms are computed in the free IR sector. Finally, recall that if we are interested in flows in which the UV fixed point is also strongly coupled (but can still be reached from a free parent theory), a more useful expression is \kUIRintfreecomp, which we reproduce below
\eqn\kUIRintfreecompi{
\tau_U^{IR}=-{27\over4}\Tr\ \hat R^{UV}_{\rm p}(\CR_{\rm vis, p}^{UV}-\hat R^{UV}_{\rm p})^2+{27\over4}\Tr|_{\rm free}\ \hat \CR^{IR}(\CR^{IR}_{\rm vis}-\hat \CR^{IR})^2+\Tr|_{\rm free}\left(U^{IR}_{\rm vis}\right)^2~.
}

\subsec{Adjoint SQCD}
In this section we will consider $SU(N_c)$ SQCD with an adjoint, $X$, $N_f$ flavors $Q$, $\tilde Q$, and no superpotential \KutasovIY. The theory has a family of $R$-symmetries given by $R^t(Q)=R^t(\tilde Q)=y, \ \ \ R^t(X)={N_f(1-y)\over N_c}$. The corresponding $U$-charges are $U^t(Q)=U^t(\tilde Q)={3\over2}y-1, \ \ \ U^t(X)=-1-{3N_f(y-1)\over 2N_c}$.

We determine $\CR_{\rm vis}$ by maximizing
\eqn\aadj{\eqalign{
a&=2(N_c^2-1)+3\left(2N_cN_f(y-1)^3+(N_c^2-1)\left({N_f(1-y)\over N_c}-1\right)^3\right)\cr&-\left(2N_cN_f(y-1)+(N_c^2-1)\left({N_f(1-y)\over N_c}-1\right)\right)~.
}}
This procedure yields
\eqn\Rvis{
y_{\rm vis}=-{-6N_c^4+3N_cN_f-3N_c^3N_f-3N_f^2+3N_c^2N_f^2+N_c\sqrt{20N_c^6+N_f^2-N_c^4(16+N_f^2)}\over3(2N_c^4+N_f^2-N_c^2N_f^2)}~,
}
and so the UV flavor two point function coefficient is
\eqn\kUUVadj{
\tau_U^{UV}={1\over2}N_cN_f(2-3y_{\rm vis})^2+(N_c^2-1)\left(1+{3N_f(y_{\rm vis}-1)\over 2N_c}\right)^2~.
}

In order to compute the IR flavor two point function coefficient \kUIRintfree\ we must identify the IR superconformal $R$ symmetry. Therefore, we must keep track of the $\CR_{\rm vis}$ charges of the various gauge invariant operators
\eqn\ginvadjSQCD{
M_j=QX^{j-1}\tilde Q, \ \ \ \Tr X^j, \ \ \ B^{(n_1,...,n_k)}=Q^{n_1}_{(1)}...Q^{n_k}_{(k)}, \ \ \ \sum_{\ell=1}^kn_{\ell}=N_c, \ \ \ k\ge1~,
}
where $Q_{(\ell)}=X^{\ell-1}Q$ are the \lq\lq dressed" quarks.

 In general this is tedious, so we first specialize to the case that $N_c=3$ while working in the asymptotically free regime, $N_f<6$. It is straightforward to check that the $B$'s and $M_j$'s never hit the unitarity bound in this regime. However, as we vary $N_f$, some of the $\Tr X^j$ will drop below the unitarity bound.

\bigskip 
\centerline{\bf $N_c=3$, $2\le N_f<6$}~
In this regime the dimensions of $\Tr X^j$ are above the unitarity bounds and so the theory has no accidental symmetries in the IR. Therefore
\eqn\kconjadj{
\tau_U^{UV}>0=\tau_U^{IR},
}
as desired.
\bigskip
\centerline{\bf $N_c=3$, $N_f=1$}~
In this case, the operators $\Tr X^2$ and $\Tr X^3$ both fall below the unitarity bound. The deformed $a$ is then
\eqn\adefadj{
a\to a+{4\over9}y^2(3+2y)+{1\over9}(-1+3y)^2(2+3y)~.
}
Maximizing this expression, we find $\tilde y={1\over63}(58-\sqrt{907})$.
Therefore, 
\eqn\kIRadj{
\tau_U^{IR}\sim.16
}
On the other hand,
\eqn\kUVadjex{
\tau_U^{UV}\sim4.87>\tau_U^{IR}.
}

\bigskip
\centerline{\it Relevant deformations}~
An additional nontrivial test of our conjecture is to start at the interacting $SU(3)$ adjoint SQCD fixed point with $N_f=5,...,2$, compute $\tau_U^{UV}$,  flow down to the fixed point with $N_f=1$, and finally compare to \kIRadj. To that end, consider deforming the $N_f=5$ fixed point by the following relevant operator
\eqn\Nfiiflow{
W=\lambda\left(Q_1\tilde Q^1+Q_2\tilde Q^2+...Q_4\tilde Q^4\right)~.
}
Using \kUIRintfreecompi\ and recalling \seconddem, we find
\eqn\tauUVNfvadj{
\tau_U^{UV}\sim12.10>.16\sim \tau_U^{IR}~.
}

Starting instead at the $N_f=4$ fixed point and turning on $W=\lambda\left(Q_1\tilde Q^1+Q_2\tilde Q^2+Q_3\tilde Q^3\right)$, we find
\eqn\tauUVNfivadj{
\tau_U^{UV}\sim11.21>.16\sim \tau_U^{IR}~.
}
Similarly, for the $N_f=3$ fixed point, we have
\eqn\tauUVNfiiiadj{
\tau_U^{UV}\sim9.45>.16\sim \tau_U^{IR}~.
}
Finally, for the $N_f=2$ fixed point, we have
\eqn\tauUVNfiiiadj{
\tau_U^{UV}\sim6.16>.16\sim \tau_U^{IR}~.
}

\bigskip
\centerline{\it The Veneziano Limit}~
Finally, let us consider adjoint SQCD in the limit of $N_c, N_f\gg1$ with $x=N_c/N_f$ fixed. All of our expressions are to leading order in $N_f$ in this limit. In particular, we find
\eqn\avenqdj{
a={N_f^2(y-1)\over x}(-9x(y-1)-3(y-1)^2+2x^2(-2-6y+3y^2))~,
}
while
\eqn\yvisvenadj{
y_{\rm vis}={3-3x-6x^2+x\sqrt{20x^2-1}\over3-6x^2}~.
}

It is easy to see that the baryons never hit the unitarity bound in this limit. The mesons $M_j$ and the $\Tr X^j$ operators do, however, drop below the unitarity bound as we increase $x$. Since each $M_j$ contains $N_f^2\gg 1$ degrees of freedom, it turns out that their contribution to $a$ and hence to the various currents we are interested in dominates the contribution from the $\Tr X^j$ operators.\foot{This statement is true as long as $x>3+\sqrt{7}$, when $M_1$ becomes free. For $1/2<x<3+\sqrt{7}$, the $\Tr X^j$ for $j=2,...,5$ sequentially decouple. However, it is straightforward to show that $\tau_U^{UV}>\tau_U^{IR}$ in this case since $\tau_U^{UV}$ scales as $\CO(N_c^2, N_cN_f)$ while $\tau_U^{IR}$ is sub-leading.}

For $x>3+\sqrt{7}$, $y_{\rm vis}<1/3$ and so $M_1=Q\tilde Q$ falls below the unitarity bound and
\eqn\adefMzadj{
a\to a^{(1)}=a+{1\over9}N_f^2(2-6y)^2(5-6y)~.
}
Therefore
\eqn\ytilde{
\tilde y^{(1)}={3+9x-6x^2+\sqrt{x(-16+87x-48x^2+20x^3)}\over3+24x-6x^2}~,
}
determines the superconformal $R$-charge for $3+\sqrt{7}<x<9.95$ (until $M_2$ decouples). In this regime, it is straightforward to check that
\eqn\kconjadj{
\tau_U^{UV}>\tau_U^{IR}~,
}
as desired.

Finally, let us check the limit $x\gg1$. First, we define $\tilde y^{(p)}$ to be the superconformal $R$-charge for $Q, \tilde Q$ after $M_p$ hits the unitarity bound but before $M_{p+1}$ does, and we let $x^{(p)}$ be the value of $x$ at which $M_{p+1}$ hits the unitarity bound. By maximizing
\eqn\apadj{
a^{(p)}=a+{1\over9}N_f^2\sum_{j=1}^p\left(2-3\left(2y+(j-1){(1-y)\over x}\right)\right)^2\left(5-3\left(2y+(j-1){(1-y)\over x}\right)\right)~,
}
we find $y^{(p)}(x)$ (we do not give the precise form here because it is too complicated). Solving
\eqn\unitbdadj{
2y^{(p)}(x)+p{1-y^{(p)}(x)\over x}={2\over3},
}
we find $x^{(p)}$ (again, we do not give the precise form of $x^{(p)}$ here). Taking the limit of large $p$, we find
\eqn\largepadj{\eqalign{
\lim_{p\to\infty}x^{(p)}&=p\left({5\over2}+\sqrt{3}\right)+{1\over13}\left(41+20\sqrt{3}\right)+\CO(p^{-1})\cr\lim_{p\to\infty}y^{(p)}&={\sqrt{3}-1\over3}+\CO(p^{-1}),\cr \lim_{p\to\infty}y_{\rm vis}&={3-\sqrt{5}\over3}+\CO(p^{-1})~,
}}
where we take $x=x^{(p)}$ at large $x$. In this limit, therefore
\eqn\kIRadjlx{
\tau_U^{IR}=N_f^2p\left(12-16\sqrt{3}+\sqrt{485-{275\sqrt{3}\over2}}\right)~,
}
while
\eqn\kUVadjlx{
\tau_U^{UV}=N_f^2p^2\left({5\over2}+\sqrt{3}\right)^2~.
}
Therefore,
\eqn\kconjlargexadj{
\tau_U^{UV}>\tau_{U}^{IR}~.
}
This confirms our conjecture in adjoint SQCD in the limit of $N_c\gg N_f\gg1$. The heuristic reason this inequality holds is simple: the free adjoint field contribution dominates all other contributions.

\bigskip
\centerline{\it Relevant deformations}~
We can also consider starting from fixed points in the Veneziano limit with $x<3+\sqrt{7}$ and turning on deformations $\lambda Q_i\tilde Q^i$ for $i=1,..., k$ with $k>N_f\left(1-{x\over3+\sqrt{7}}\right)$ (but keeping $k$ small enough so that $QX\tilde Q$ doesn't decouple). We then flow to a theory with $N_f^2$ free $Q\tilde Q$ mesons. Although the expressions are complicated, it is straightforward to use \kUIRintfree\ and find that
\eqn\tauvenadiint{
\tau_U^{UV}>\tau_U^{IR}~,
}
as desired.

\bigskip
\centerline{\it Higgsing}~
Another nontrivial test of our conjecture is to consider Higgsing the above theory in the Veneziano limit with $1/2<x<\sqrt{5/2}$ (i.e., before any of the singlets decouple in the IR). We consider turning on an expectation value for the adjoint $\langle X\rangle={\rm diag}(x_1^{Nc/l}, x_2^{Nc/l},..., x_l^{N_c/l})$, where $x_i^{N_c/l}$ means a block of $N_c/l$ consecutive $x_i$'s on the diagonal (we take all the $x_i\ne0$ to be distinct and take $l\ll N_f, N_c$ for simplicity). This expectation value Higgses $SU(N_c)\to SU(N_c/l)^l\times U(1)^{l-1}$ and leaves a set of $l$ decoupled adjoint SQCD sectors with $N_c/l$ colors and $N_f$ flavors of $Q, \tilde Q$ (there are also $l-1$ singlets under $SU(N_c/l)^l$, $\hat X$, which will play a sub-leading role in what follows). We compute the $R_{\rm vis}$ current by demanding that it leave the vacuum invariant and recalling \seconddem. We then find $\CR_{\rm vis}(X)=0, \ \CR_{\rm vis}(Q)=\CR_{\rm vis}(\tilde Q)=1$.

At the interacting adjoint SQCD fixed point, we use anomaly matching and find that
\eqn\tauUVadjSQCDhiggs{
\tau_U^{UV}={x^2N_f^2\left(3+4\sqrt{20x^2-1}+10x^2\left(-6+\sqrt{20x^2-1}\right)\right)\over 2(1-2x^2)^2}~.
}
We will imagine that $b\equiv x/l\le{\rm min}\left(1/2, x/2\right)$ so that the decoupled adjoint SQCD theories are IR free. Therefore, at long distances, we find
\eqn\tauIRadjSQCDhiggs{
\tau_U^{IR}={N_f^2\over2}x(1+2b)~.
}
It is straightforward to check that for $b\le{\rm min}\left(1/2,x/2\right)$ (the regime of validity of \tauIRadjSQCDhiggs)
\eqn\tauconjhiggsasqcd{
\tau_U^{UV}>\tau_U^{IR}~.
}

\subsec{The $\hat D$ SCFT}
In this section we consider $SU(N_c)$ SQCD with two adjoints, $X, Y$, and $N_f$ fundamental flavors, $Q, \tilde Q$  and the following superpotential 
\eqn\DW{
W=\Tr XY^2~.
}
In the deep IR, this theory flows to an interacting SCFT called the $\hat D$ SCFT \IntriligatorMI. The trial $R$-charges are $R^t(Q)=R^t(\tilde Q)=y, \ R^t(X)={2(1-y)\over x}, \ R^t(Y)=1-{1-y\over x}$, where $x\equiv N_c/N_f>1$ to ensure asymptotic freedom. The trial $U$-charges are $U^t(Q)=U^t(\tilde Q)={3\over2}y-1, \ U^t(X)=-1+{3(1-y)\over x}, \  U^t(Y)={-3+x+3y\over 2x}$.

We determine $\CR_{\rm vis}$ by maximizing
\eqn\aDhat{\eqalign{
a&=2(x^2N_f^2-1)+3\Big(2xN_f^2(y-1)^3+(x^2N_f^2-1)\Big(-1+{2(1-y)\over x}\Big)^3+(x^2N_f^2-1)\Big({-1+y\over x}\Big)^3\Big)\cr&-\Big(2xN_f^2(y-1)+(x^2N_f^2-1)\Big(-1+{2(1-y)\over x}\Big)+(x^2N_f^2-1)\Big({-1+y\over x}\Big)\Big)~,
}}
we find
\eqn\yvisDhat{\eqalign{
y_{\rm vis}&={1\over3(7+N_f^2x^2(-7+2x^2))}\Big(21-21N_f^2x^2+12N_f^2x^3+6N_f^2x^4\cr&-x\left(12+\sqrt{25-2N_f^2x^2(18+17x^2)+N_f^4x^4(11+38x^2)}\right)\Big)~.
}}
The UV flavor two point function coefficient is then
\eqn\kUVDhat{
\tau_U^{UV}={1\over2}xN_f^2\left(2-3y_{\rm vis}\right)^2+(x^2N_f^2-1)\left(-1+{3(1-y_{\rm vis})\over x}\right)^2+(x^2N_f^2-1)\left({-3+x+3y_{\rm vis}\over 2x}\right)^2~.
}
To compute the IR superconformal $R$ symmetry we must again keep track of the $\CR_{\rm vis}$ charges of the various gauge invariant operators \IntriligatorMI
\eqn\ginvDhat{
\Tr X^{\ell} \ (\ell\ge2),  \ \ M_{\ell,j}=\tilde QX^{\ell}Y^jQ \ (\ell \ge0, j=0,1),  \ \ Q_{(0,0)}^{n_{(0,0)}}Q_{(1,0)}^{n_{(1,0)}}...Q_{(0\ell,0)}^{n_{(\ell,0)}}Q_{(0,1)}^{n_{(0,1)}}Q_{(1,1)}^{n_{(1,1)}}...Q_{(0,0)}^{n_{(k,1)}}~,
}
where 
\eqn\defQDhat{
Q_{(\ell, j)}=X^{\ell}Y^jQ, \ \ \ \ell\ge0, \ j=0,1~,
}
and
\eqn\sumcond{
\sum_{j=0}^{\ell}n_{(j,0)}+\sum_{j=0}^kn_{(k,1)}=N_c~.
}
Other chiral gauge invariant operators are related to the operators in \ginvDhat\ by the chiral ring relations following from the superpotential in \DW. 

A general analysis of this theory is again tedious and so we first specialize to the case $N_c=4$ while working in the asymptotically free regime, $N_f<4$. The baryons and the $\Tr X^{\ell}$ operators never hit the unitarity bound in this theory. However, as we lower $N_f$, one of the mesons does fall below the unitarity bound.

\bigskip
\centerline{$N_c=4$, $1<N_f<4$}~
One can check that in this case none of the operators hit the unitarity bound and so
\eqn\unitbdNciiiDhat{
\tau_U^{UV}>0=\tau_U^{IR}~,
}
as desired.

\bigskip
\centerline{$N_c=4$, $N_f=1$}~
In this case, we have
\eqn\yvisNcivNfi{
y_{\rm vis}\sim.32424~,
}
and so $M_{(0,0)}=Q\tilde Q$ falls below the unitarity bound. The deformed $a$ is then
\eqn\defaDhatNciv{
a\to a+{1\over9}(2-6y)^2(5-6y)~.
}
Maximizing this expression, we find $\tilde y\sim.32411$. Therefore,
\eqn\kUIRivDhat{
\tau_U^{IR}\sim.0007~.
}
On the other hand
\eqn\kUUVivDhat{
\tau_U^{UV}\sim6.67>\tau_U^{IR}~.
}

\bigskip
\centerline{\it Relevant deformations}~
Another nontrivial test of our conjecture is to start at the interacting $SU(4)$ $\hat D$ fixed point with $N_f=3,2$, compute $\tau_U^{UV}$,  flow down to the fixed point with $N_f=1$, and finally compare to \kUIRivDhat. To that end, deforming the $N_f=3$ fixed point by a relevant operator
\eqn\Nfiiflow{
W=\lambda\left(Q_1\tilde Q^1+Q_2\tilde Q^2\right)~,
}
and using \kUIRintfree, \seconddem, we find
\eqn\tauUVNfiii{
\tau_U^{UV}\sim12.09>.0007\sim \tau_U^{IR}~.
}

Starting instead at the $N_f=2$ fixed point and turning on a deformation of the form
\eqn\Nfiiflow{
W=\lambda Q_1\tilde Q^1~,
}
we find
\eqn\tauUVNfii{
\tau_U^{UV}\sim9.42>.0007\sim\tau_U^{IR}~.
}

\bigskip
\centerline{\it The Veneziano limit}~
Here we consider our theory in the limit of $N_c, N_f\gg1$ with $x$ fixed. In this limit,
\eqn\avenDhat{
a=N_f^2{y-1\over x}(-36x(y-1)-21(y-1)^2+x^2(-13-12y+6y^2))~,
}
and
\eqn\yvisvenDhat{
y_{\rm vis}=-{21-6x(2+x)+x\sqrt{11+38x^2}\over 6x^2-21}~.
}
It is not difficult to see that no baryons hit the unitarity bound in this limit. The mesons $M_{(\ell,j)}$ and the $\Tr X^j$ operators do, however, hit the unitarity bounds as $x$ increases. Again, since the mesons contain $N_f^2\gg1$ degrees of freedom, their contribution to the currents we wish to study dominate the contributions from the $\Tr X^j$ operators.\foot{Unlike in the case of adjoint SQCD, the first operator to hit the unitarity bound is $M_{(1,0)}$.}

For $x>{2\over11}(12+\sqrt{67})$, $y_{\rm vis}<1/3$ and so $M_{(1,0)}=Q\tilde Q$ falls below the unitarity bound. Therefore, we must deform $a$ as follows
\eqn\adefDhat{
a\to a^{(1)}=a+{1\over9}N_f^2(2-6y)^2(5-6y)~,
}
and we find that
\eqn\ytildeDhat{
\tilde y^{(1)}={-21+6x^2-\sqrt{-112x+315x^2-120x^3+38x^4}\over3(-7-8x+2x^2)}~,
}
determines the superconformal $R$-charge for ${2\over11}(12+\sqrt{67})<x<6.135$ (until $M_{(2,0)}$ decouples). It is then straightforward to verify that
\eqn\kconjvenDhat{
\tau_U^{UV}>\tau_U^{IR}~,
}
as desired.

We can again check the limit $x\gg1$ as in the case of adjoint SQCD in the previous section. We again define $\tilde y^{(p)}$ and $x^{(p)}$ as in the previous section (note that operators that include $Y$ never violate the unitarity bound). Therefore, we must maximize 
\eqn\apadj{
a^{(p)}=a+{1\over9}N_f^2\sum_{j=1}^p\left(2-3\left(2y+2(j-1){(1-y)\over x}\right)\right)^2\left(5-3\left(2y+2(j-1){(1-y)\over x}\right)\right)~.
}
Then, we find $y^{(p)}(x)$ (we do not give the precise form here because it is too complicated). Solving 
\eqn\unitbdadj{
2y^{(p)}(x)+2p{1-y^{(p)}(x)\over x}={2\over3},
}
we find $x^{(p)}$ (again, we do not give the precise form of $x^{(p)}$ here). Taking the limit of large $p$, we find
\eqn\largepadj{\eqalign{
\lim_{p\to\infty}x^{(p)}&={27\over11}p+\CO(p^{-1})\cr\lim_{p\to\infty}y^{(p)}&=-{1\over8}+\CO(p^{-1}),\cr \lim_{p\to\infty}y_{\rm vis}&={6-\sqrt{38}\over6}+\CO(p^{-1})~,
}}
where we take $x=x^{(p)}$ at large $x$. In this limit, therefore
\eqn\kUVfinalDhat{
\tau_U^{UV}={3645\over484}N_f^2p^2+\CO(p)
}
On the other hand
\eqn\kIRfinalDhat{
\tau_U^{IR}={11(-5109635+835272\sqrt{38})\over2239488}N_f^2p
}
Since $p\gg1$, it follows that
\eqn\kconjlargexDhat{
\tau_U^{UV}>\tau_U^{IR}~,
}
is satisfied in this regime. The reason is again simple: the contributions from the $\sim 2N_c^2$ degrees of freedom in the free UV adjoints dominate all other contributions to the two point function coefficients.

\bigskip
\centerline{\it Relevant deformations}~
We can also consider starting from fixed points in the Veneziano limit with $x<{2\over11}(12+\sqrt{67})$ and turning on deformations $\lambda Q_i\tilde Q^i$ for $i=1,..., k$  with $k>N_f\left(1-{11x\over2(12+\sqrt{67})}\right)$ (but keeping $k$ small enough so that $QX\tilde Q$ doesn't decouple). We then flow to a theory with $N_f^2$ free $Q\tilde Q$ mesons. Although the expressions are complicated, it is straightforward to use \kUIRintfree\ and find that
\eqn\tauvenadiint{
\tau_U^{UV}>\tau_U^{IR}~,
}
as desired.

\bigskip
\centerline{\it Higgsing}~
In this section, we consider Higgsing the above theory in the Veneziano limit, with $1<x<{2\over11}(12+\sqrt{67})$. We will consider turning on the vevs, $\langle X\rangle={\rm diag}(x_1^{Nc/l}, x_2^{Nc/l},..., x_l^{N_c/l})$, where $x_i^{N_c/l}$ means a block of $N_c/l$ consecutive $x_i$'s on the diagonal (we choose all $x_i\ne0$ and distinct, and we take $l\ll N_f, N_c$ for simplicity). This procedure Higgses the gauge group $SU(N_c)\to SU(N_c/l)^l\times U(1)^{l-1}$. Each non-abelian sub-sector has $N_f$ fundamental flavors $Q$, $\tilde Q$, and an adjoint, $X$ (the corresponding $Y$ adjoint is rendered massive by the superpotential \DW). For $l=2$, the $Y$ field gives rise to massless bifundamentals $Y_{1,2}$ and $Y_{2,1}$ transforming in the ${\bf N_c/2\times\bar{N_c/2}}$ and ${\bf \bar{N_c/2}\times{N_c/2}}$ representations of $SU(N_c/2)^2$ respectively. For $l>2$ these fields are generally rendered massive by the superpotential in \DW\ (consider for example turning on the vevs $x_a=a$ for $a=1,...,l-1$ and $x_l=-l(l-1)/2$). Note that the theory has an additional $l-1$ singlets $\hat X$ coming from the original $X$ adjoint whose effects are sub-leading in the limit we study (with the vevs we have chosen, the corresponding $\hat Y$ singlets from the original $Y$ adjoint are all massive).

We compute $\CR_{\rm vis}$ by demanding that it leave the vacuum invariant and recalling the discussion around \seconddem
\eqn\RvisDhathiggs{
\CR_{\rm vis}(X)=0, \ \ \ \CR_{\rm vis}(Y)=1, \ \ \ \CR_{\rm vis}(Q)=\CR_{\rm vis}(\tilde Q)=1~.
}
At the interacting UV fixed point, we use anomaly matching and find
\eqn\tauUVDhathiggs{
\tau_U^{UV}=-{N_f^2x\left(264x+912x^3-x(155+38x^2)\sqrt{11+38x^2}\right)\over 4(7-2x^2)^2}~.
}
In the deep IR, taking $l>2$ and assuming that we Higgs enough of the UV gauge group so that the IR is free, we find
\eqn\tauIRDhathiggs{
\tau_U^{IR}=N_f^2\left(bx+{x\over2}\right)~,
}
where $b\equiv x/l$. It is then easy to check that for $b\le1/2$ (the range of validity of the above expression)
\eqn\tauconjhiggsDhat{
\tau_U^{UV}>\tau_U^{IR}~,
}
as desired. 

Finally, consider the case that $l=2$. In this case, we must include the $Y_{1,2}$ and $Y_{2,1}$ bifundamentals in the IR. We find
\eqn\tIRbifund{
\tau_U^{IR}={N_f^2\over2}\left({3x^2\over2}+x\right)~,
}
and so once again
\eqn\tauconjDhatiii{
\tau_U^{UV}>\tau_U^{IR}~.
}

\appendix{G}{Misleading anomaly matching: $SO(N)$ with an ${\bf N(N+1)/2-1}$}
In this appendix, we consider the Brodie, Cho, and Intriligator theory \BrodieVV: an $SO(N)$ gauge theory (taking $N\ge5$ to ensure asymptotic freedom) with a traceless symmetric tensor, $S$. This theory is an example of a misleading anomaly matching because $S$ has $R$-charge $\CR_{\rm vis}^{UV}(S)={4\over N+2}$,
while the chiral gauge invariant composites, $\CO_n=\Tr S^n$, have $R$-charge $\CR_{\rm vis}^{IR}(\CO_n)={4n\over N+2}$, and so the $R$-anomalies of the UV and (putative) IR theories match. However, it can be shown that the IR is actually in an interacting phase with only a subset of the $\CO_n$ decoupling \BrodieVV.

In fact, $\tau_U$ is sensitive to the IR phase of this theory and gives us another opportunity to check our conjecture. To that end, note that $U(S)=-1+{6\over N+2}, \ \ \ U(\CO_n)=-1+{6n\over N+2}$, and so 
\eqn\tauuvBCI{
\tau_U^{UV}={(N-4)^2(N-1)\over2(N+2)}, \ \ \ \tau_U^{IR, \rm confining}={7N^3+3N^2+6N-16\over(N+2)^2}~.
}
It is easy to see that $\tau_U^{UV}< \tau_U^{IR, \rm confining}$ for $4<N<21$. We conclude that, at least in this range of $N$, the confining description is inconsistent with our conjecture. While $\tau_U^{UV}>\tau_U^{IR, \rm confining}$ for $N>20$, this does not contradict our conjecture---it only means that our criterion cannot give us more information for this range of $N$.

We can take into account the unitarity bound in our computations and note that while the fields with $n\le(2+N)/6$ must decouple, the theory can still include an interacting module. If we assume that no other fields decouple, we find
\eqn\tauIRBCI{
\tau_U^{IR}={N^3-21N^2+138N-280\over18(N+2)^2}~.
}
It is then easy to check that
\eqn\tauconjBCI{
\tau_U^{UV}>\tau_U^{IR}~.
}

\listrefs
\end